\documentclass[11pt,a4paper,onecolumn]{article}

\usepackage{graphicx}
\usepackage{amssymb}

\begin{document}


\input epsf
\epsfverbosetrue

\setcounter{page}{1}
\pagenumbering{arabic}

\vspace{66pt}
\begin{center}
{\bf
EXTENSION OF THE CEM2k and LAQGSM CODES TO DESCRIBE PHOTO-NUCLEAR
REACTIONS
}\\
\vspace{33pt}
{\bf
S. G. Mashnik$^{1}$,
M. I. Baznat$^{2}$,
K. K. Gudima$^{2}$,
A. J. Sierk$^{1}$,
R. E. Prael$^{1}$
}

\vspace{-1mm}
$^1$Los Alamos National Laboratory, Los Alamos, New Mexico 87545, USA\\
$^{2}$Institute of Applied Physics, Academy of Science of
Moldova, Chi\c{s}in\u{a}u, Moldova

\end{center}
\vspace{33pt}
\begin{center}
{\bf Abstract}
\end{center}
The improved Cascade-Exciton Model (CEM) code CEM2k+GEM2 and the Los Alamos 
version of the Quark-Gluon String Model code LAQGSM are extended 
to describe photonuclear reactions. First, we incorporate into 
CEM2k+GEM2 new evaluations of elementary cross sections based on the
latest experimental data and also make several improvements in
the description of the de-excitation of nuclei remaining 
after the cascade stage of reactions induced by arbitrary 
projectiles. Next, for photonuclear reactions we include 
in CEM2k+GEM2 a normalization to evaluated experimental absorption 
cross sections based on 
the recent systematics by Kossov.
Then, we extend our high-energy code LAQGSM by adding the 
photonuclear mode which was ignored in all its
previous versions, and add to it the photonuclear
part from our improved CEM2k+GEM2.
In this work we present a short
description of the photonuclear mode as incorporated into our
codes, show several illustrative results, and 
point out some unresolved problems.

\newpage
\noindent{\bf Introduction}

\indent
The 2003 version \cite{OurNewINC,Madeira04,GSI03}
of the improved \cite{CEM2k} Cascade-Exciton Model 
(CEM) \cite{CEM} as realized in the code CEM2k
merged \cite{Mashnik02a,Mashnik02c,fitaf}
with the Generalized Evaporation Model code GEM2 by Furihata
\cite{GEM2} and of the Los Alamos version of the Quark-Gluon
String Model code LAQGSM \cite{LAQGSM}
merged \cite{Mashnik02a,Mashnik02c,fitaf}
with GEM2 have been recently incorporated 
into the transport codes MARS15 \cite{MARS15} and LAHET3 
\cite{LAHET3} and are planned to be incorporated in the 
future into the transport codes MCNPX \cite{MCNPX}
and MCNP6 \cite{MCNP6}. Initially, neither CEM2k+GEM2
nor LAQGSM+GEM2 considered photonuclear reactions and were
not able to describe such reactions, either as 
stand-alone codes or as event generators in transport codes.
To address this problem, here we extend
CEM03 (the 2003 version of CEM2k+GEM2) and LAQGSM03
(the 2003 version of LAQGSM+GEM2) codes to describe photonuclear reactions 
at intermediate energies (from $\sim 30$ MeV to $\sim 1.5$ GeV). 
We develop a model that is based on the
Dubna IntraNuclear Cascade (INC)
Photonuclear Reaction Model (PRM) \cite{Dubna69}--\cite{Dubna74},
uses experimental data now available in the literature, and
a revision of recent systematics for the total photoabsorption
cross sections by Kossov \cite{Kossov}. Our photonuclear reaction
model still has some problems and is under further development,
but even the current version allows us to describe
reasonably well
intermediate energy photonuclear reactions. In the following, we 
present a description of our model together with several
illustrative results. 

\noindent{\bf Dubna Photonuclear Reaction Model}

\indent
The Dubna intranuclear cascade photonuclear reaction model
(Dubna INC) was initially developed 35 years ago by
one of us (KKG) in collaboration with Iljinov and Toneev \cite{Dubna69}
to describe photonuclear reactions at energies
above the Giant Dipole Resonance (GDR) region.
[At photon energies $T_{\gamma} = 10$--$40$ MeV, the de Broglie 
wavelength $\lambda$ is of the order of $20$--$5$ fm,  
greater than the average inter-nucleonic distance in the nucleus; 
the photons interact with the nuclear
dipole resonance as a whole, thus the INC is not applicable.]
Below the pion production threshold, the Dubna INC considers
absorption of photons on only ``quasi-deuteron" pairs according
to the Levinger model \cite{Levinger}:
\begin{equation}
\sigma_{\gamma A} = L \frac{Z(A-Z)}{A} \sigma_{\gamma d} \mbox{ ,} 
\end{equation}
where $A$ and $Z$ are the mass and charge numbers of the nucleus,
$L \approx 10$, and $ \sigma_{\gamma d}$ is 
the total photoabsorption cross section on deuterons as
defined from experimental data.

At photon energies above the pion-production threshold, the Dubna INC
considers production of 
one or two pions; the concrete
mode of the reaction is chosen by the Monte Carlo method according to the
partial cross sections (defined from available experimental data):
\begin{eqnarray}
\gamma + p & \to & p + \pi^0  \mbox{ ,} \\ 
           & \to & n + \pi^+ \mbox{ ,}  \\
           & \to & p + \pi^+ + \pi^-  \mbox{ ,}  \\
           & \to & p + \pi^0 + \pi^0  \mbox{ ,}  \\
           & \to & n + \pi^+ + \pi^0  \mbox{ .} 
\end{eqnarray}
The cross sections of $\gamma + n$ interactions are derived from 
consideration of isotopic invariance, {\it i.e.,} it is assumed that
$\sigma (\gamma + n) = \sigma (\gamma + p)$. The Compton effect on 
intranuclear nucleons is neglected, as its cross section is less
than $\approx 2$\% of other reaction modes (see, {\it e.g.,}
Fig.\ 6.13 in Ref.\ \cite{LockMeasday}). The Dubna INC does not
consider processes
involving production of three and more pions; this limits the model
applicability to photon energies $T_{\gamma} \lesssim 1.5$ GeV
[for $T_\gamma$ higher than the threshold for three-pion production,
the sum of the cross sections (4)--(6) is assumed to be equal to 
the difference
between the total inelastic $\gamma + p$ cross section 
and the sun of the cross sections of the two-body reactions (2)--(3)].

The kinematics of two-body reactions (2)--(3) and absorption of photons 
by a pair of nucleons is completely defined by a given direction of
emission of one of the secondary particles. Similarly to the procedure 
followed for $N+N$ and $\pi + N$ interactions 
\cite{INC68a,INC68b}, 
the cosine of the angle 
of emission of secondary particles can be represented in the c.m. system
as a function of a random number $\xi$, distributed uniformly in the interval
[0,1]
\begin{equation}
\cos \theta = 2 \xi ^{1 / 2} \left[ \sum_{n=0}^{N} a_n \xi^n
+ (1- \sum_{n=0}^N a_n ) \xi^{N+1} \right] -1 \mbox{ ,}
\end{equation}
where $N = M = 3$,
\begin{equation}
a_n = \sum_{k=0}^M a_{nk} T_{\gamma}^k \mbox{ ,}
\end{equation}
where the coefficients $a_{nk}$ were fitted to describe the available 
experimental data and are published in Tabs.\ 1 and 2 of Ref.\ \cite{Dubna74}
(the corresponding coefficients for $N+N$ and $\pi + N$ interactions
are published in Tab.\ 3 of Ref.\ \cite{INC68b} and in
Tab.\ 72 of the monograph \cite{Book}).
The distribution of the secondary particles over the azimuthal angle
$\varphi$ is assumed to be isotropic. After simulating the angle
$\Theta_1$, using Eqs.\ (7--8), and $\varphi_1$ isotropically for the first
particle of any reaction with two particles in the final state,
the angles $\Theta_2$ and  $\varphi_2$ of the second particle, as well
as the energies of both particles $T_1$ and $T_2$ are uniquely determined
from four-momentum conservation.

The analysis of experimental data has shown that the channel (4)
of two-pion photoproduction proceeds mainly through the decay of the 
$\Delta^{++}$ isobar listed in the last Review of Particle Physics by the
Particle Data Group \cite{RPP04} as having the mass $M = 1232$ MeV
\begin{eqnarray}
\gamma + p  & \to & \Delta^{++} + \pi^-  \mbox{ ,} \nonumber \\
\Delta^{++} & \to & p + \pi^+  \mbox{ ,} 
\end{eqnarray}
whereas the production cross section of other isobar components
$\left({3\over 2}, {3\over 2} \right)$ are small and can be neglected.
The Dubna INC uses the Lindenbaum-Sternheimer resonance model \cite{Delta}
to simulate the reaction (9).
In accordance with this model, the mass of the isobar $M$ is determined from
the distribution
\begin{equation}
{{\mathrm{d} W}\over {\mathrm{d} M}} \sim F(E,M) \sigma(M) \mbox{ ,}
\end{equation}
where $E$ is the total energy of the system, $F$ is the two-body phase
space of the isobar and $\pi^-$ meson, and $\sigma$ is the 
isobar production cross section which is assumed to be equal to the
cross section for elastic $\pi^+ p$ scattering.

The c.m. emission angle of the isobar is approximated
using Eqs.\ (7) and (8) with the coefficients $a_{nk}$ listed in Tab.\ 3 of 
Ref.\ \cite{Dubna74}; isotropy of the decay of the isobar in its c.m.\ system 
is assumed.

In order to calculate the kinematics of the non-resonant
part of the reaction (4) and two remaining three-body channels (5) and (6),
the Dubna INC uses the statistical model. The total energies of the two 
particles 
(pions) in the c.m.\ system are determined from the distribution
\begin{equation}
{ {\mathrm{d} W}\over {\mathrm{d} E_{\pi_1} \mathrm{d} E_{\pi_2}} }
\sim (E-E_{\pi_1}-E_{\pi_2}) E_{\pi_1}E_{\pi_2} / E \mbox{ ,}
\end{equation}
and that of the third particle (nucleon, $N$) from 
conservation of energy.
The actual simulation of such reactions is done as follows:
Using a random number $\xi$,
we simulate in the beginning the energy of the first pion using
$$E_{\pi_1} = m_{\pi_1} + \xi (E^{max}_{\pi_1} - m_{\pi_1}) ,$$
where 
$$E^{max}_{\pi_1} = [ E^2 + m^2_{\pi_1} - (m_{\pi_2} + m_N )^2 ] / 2 E . $$
Then, we simulate the energy of the second pion $E_{\pi_2}$ 
according to Eq.\ (11) using the Monte Carlo rejection method.
The energy of the nucleon is then calculated as 
$E_N = E - E_{\pi_1} - E_{\pi_2}$, checking that the ``triangle law"
for momenta
$$| p_{\pi_1} - p_{\pi_2} | \leq p_N \leq |p_{\pi_1} + p_{\pi_2}|$$
is fulfilled, otherwise this sampling is rejected and the procedure
is repeated. The angles $\Theta$ and $\varphi$ of the pions
are sampled assuming an isotropic distribution of particles in the
c.m.\ system, 
$$ \cos \Theta_{\pi_1} = 2 \xi_1 -1, \qquad
 \cos \Theta_{\pi_2} = 2 \xi_2 -1, \qquad 
\varphi_{\pi_1} = 2 \pi \xi_3,\qquad 
\varphi_{\pi_2} = 2 \pi \xi_4,$$
and the angles of the nucleon are defined from momentum
conservation, $\vec{p}_N = - (\vec{p}_{\pi_1} + \vec{p}_{\pi_2} )$.

So an interaction of a photon with a nucleons inside a nucleus
leads to two or three fast cascade particles. Depending on their
momenta and coordinates, these particles can leave the nucleus,
be absorbed, or initiate a further 
intranuclear cascade.
All the remaining details of the Dubna INC (followed by the 
evaporation/fission of excited nuclei produced after the cascade
stage of reactions) calculation are the same as for
$N+A$ and $\pi + A$ reactions and are described in detail in
the monograph \cite{Book}.

The Dubna INC PRM was used successfully
for many years as a stand-alone model
to study different aspects of photonuclear reactions and was
also incorporated without modifications
into the transport codes CASCADE \cite{CASCADE}
and GEANT4 \cite{GEANT4}, and with some improvements,
via CEM95 \cite{CEM95}, CEM97 \cite{CEM97}, and CEM2k \cite{CEM2k},
into the transport codes
MARS14 \cite{MARS14} and MCNPX \cite{MCNPX,Franz00,Franz04}, respectively.
In the middle of the 1970's, one of the authors of the initial
version of the Dubna INC PRM, Dr. A. J. Iljinov, moved from JINR,
Dubna to INR, Moscow and continued to develop further the Dubna INC
with his Moscow Group, which evolved into what is now known in the
literature as the Moscow INC model 
(see, {\it e.g.}, \cite{Iljinov94} and references therein). 
The Moscow INC model was recently extended to describe 
photonuclear reactions at energies up to 10 GeV \cite{Iljinov97}.
 
\noindent{\bf From CEM95 to CEM03}

Photonuclear reactions were not considered in the initial
version of the CEM \cite{CEM}. The Dubna PRM was incorporated \cite{CEM95photo}
first into the CEM95 \cite{CEM95} version of the CEM and used thereafter
to analyze a large number of photonuclear reactions \cite{Gereghi90}.
Later on, CEM95 was incorporated as an event generator
into the MARS14 \cite{MARS14} transport code and used in some
applications.

By early 1997, one of the authors of the CEM (SGM) moved from
JINR, Dubna to LANL, Los Alamos, and continued to develop further
with his collaborators the cascade-exciton model
for LANL needs, {\it e.g.}, as an event generator for the Los Alamos
transport code MCNPX \cite{MCNPX} and for other applications.

\noindent{\it New Approximations for $\gamma p$ Cross Sections}

The first improvements in the CEM of the photonuclear mode of the 
Dubna INC was done in the CEM97 version \cite{CEM97} of the CEM.
The improved cascade-exciton model in the code CEM97 differs from 
the older CEM95 version by incorporating new 
approximations for the elementary $NN$, $\pi N$, and
$\gamma p$ cross sections used in the cascade, using more 
precise values for nuclear masses and pairing energies, 
employing a corrected systematics for the level-density
parameters, 
adjusting the cross sections for pion absorption on quasi-deuteron 
pairs inside a nucleus, 
including the Pauli principle in the preequilibrium calculation, 
and improving the calculation of fission widths.
Implementation of significant refinements and improvements in the 
algorithms of many subroutines led to a decrease of the computing time 
by up to a factor of 6 for heavy nuclei, which is very important when 
performing simulations with transport codes.

Concerning specifically the photonuclear reactions, in CEM97 we 
developed improved approximations for the elementary $\gamma p$ 
cross sections compared with the Dubna INC PRM \cite{Dubna69}.

In the Dubna INC PRM~\cite{Dubna69} used in CEM95,
the cross sections for the free $\gamma p$
(and for $NN$ and $\pi N$) interactions 
are approximated using a special algorithm of interpolation/extrapolation
through a number of picked points, mapping as well as possible the
experimental data.
This was done very accurately by the authors of the 
Dubna INC PRM \cite{Dubna69} using all
experimental data available at that time, about 35 years ago.
Currently there are many more experimental data on cross section; 
therefore we revised the approximations of all elementary
cross sections used in CEM97 \cite{CEM97}. 
We collected all published experimental data from available sources,
then developed an improved
algorithm for approximating cross sections and developed simple
and fast approximations for elementary cross sections which fit very well
presently available experimental data not only up to 
$\sim 1.5$ GeV, where the Dubna INC PRM is assumed to be used, or up to about
5 GeV, the upper recommended energy for the present version of the CEM
for nucleon- and pion-induced reactions, 
but up to 50--100 GeV and higher, depending on availability of data. 
So far we have such approximations for 8 different 
types of $\gamma + p$ elementary cross sections and for 24 types of
reactions induced by nucleons and pions. Cross sections for other
types of interactions taken into account by CEM are calculated from
isospin considerations using the former as input. These cross sections
are used in CEM97 \cite{CEM97}, CEM2k \cite{CEM2k}, and
CEM03 \cite{OurNewINC}, and were 
incorporated recently into the latest version of our LAQGSM \cite{LAQGSM} 
code, LAQGSM03~\cite{OurNewINC,Madeira04}.

We consider this part of the CEM improvement as an independently useful
development, as our approximations are reliable, fast, and easy to
incorporate into any transport, INC, BUU, or Glauber-type model codes.
For example, our new approximations recently have been successfully 
incorporated by Nikolai Mokhov into the MARS14 \cite{MARS14}
and MARS15 \cite{MARS15} versions of the 
MARS code system at Fermilab. 

An example of 8 compiled $\gamma + p$ experimental cross sections 
together with our approximations
and the old approximations from the Dubna INC PRM used in
CEM95 is shown in Fig.\ 1. We see that
these approximations describe very well all data.
Although presently we have much more data than 35 years ago when 
the Dubna group produced their approximations used in the 
Dubna INC PRM \cite{Dubna69},       
for a number of interaction modes like the total $\gamma p$ cross
sections at energies below 1.2 GeV (where the initial Dubna INC PRM was
assumed to be used), and for such modes as $\gamma + p \to p + \pi^0$,
$\gamma + p \to n + \pi^+$, $\gamma + p \to p + \pi^+ + \pi^-$, 
and  $\gamma + d \to n + p$, at energies not too close to their
thresholds,
the original approximations also agree very well with
presently available data, in the energy region where the Dubna INC PRM was 
developed to work.  This is a partial explanation of why the old
Dubna INC \cite{Book} and the younger CEM95 \cite{CEM95} describe so well
many characteristics of different nuclear reactions. 
On the other hand, for some elementary cross sections 
like $\gamma + p \to p + 2 \pi^0$ and $\gamma + p \to n + \pi^+ + \pi^0$, 
the old approximations differ significantly from the present data, 
demonstrating the need for a better description 
of all modes of photonuclear reactions. (Similar results were obtained
in CEM97 \cite{CEM97} for hadron-hadron cross section approximations.)

The CEM97 code with these cross sections and the other mentioned
improvements was incorporated by Gallmeier \cite{Franz00}  
into the MCNPX transport code \cite{MCNPX}, allowing MCNPX to consider
for the first time interaction of intermediate-energy photons with
thick targets of practically arbitrary geometry.

\newpage

\vspace{-5mm}
\begin{figure}[!ht]
\begin{center}
\includegraphics[width=14.0cm]{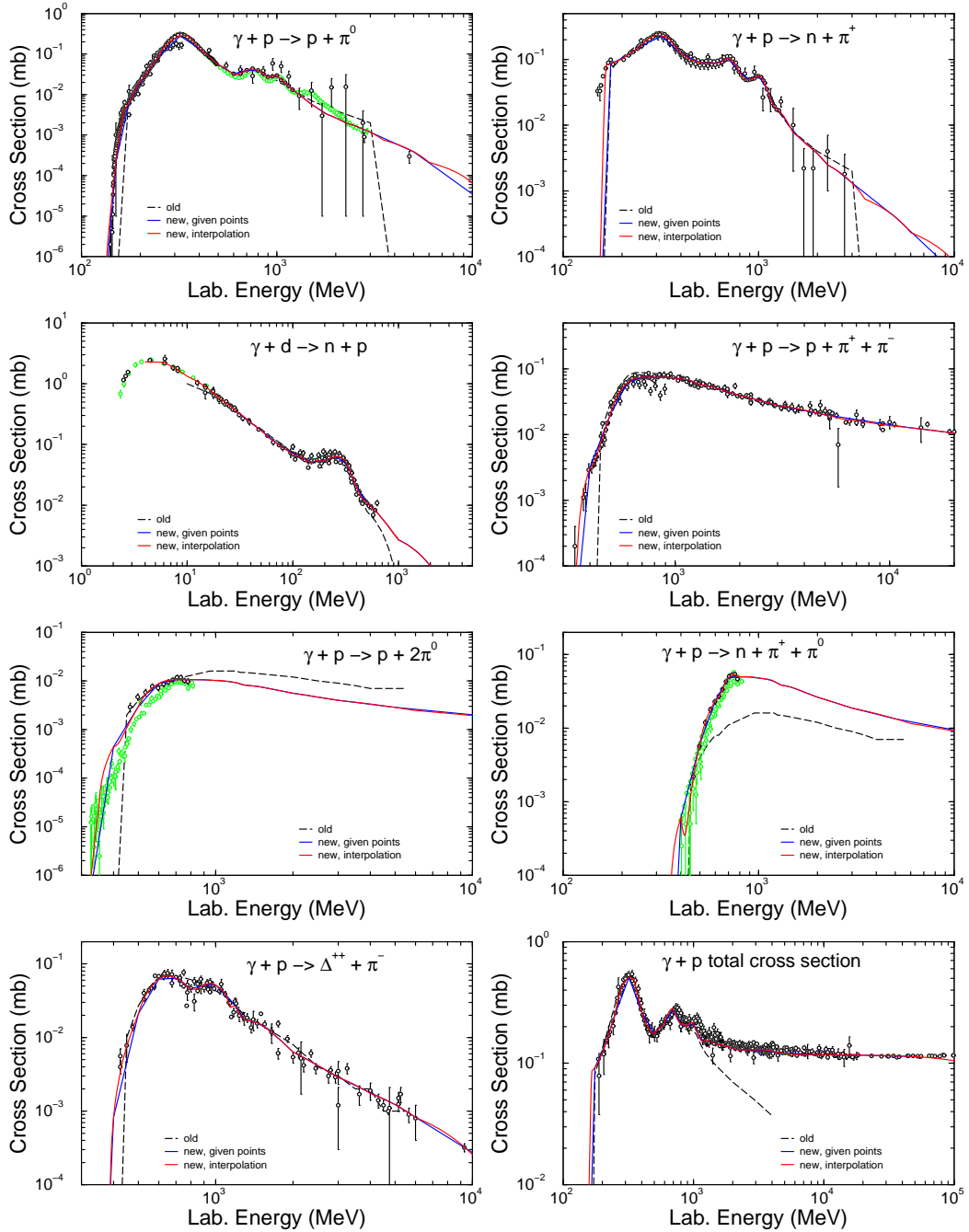}
\vspace{-2mm}
\caption{
Comparison of eight experimental total $\gamma + p(d)$ cross sections
with the old approximations used in the Dubna INC PRM and with
our approximations incorporated into the 
CEM03 and LAQGSM03 codes. The red curve gives the code results
using parabolic interpolation, while the blue solid curve uses linear 
interpolation between our tabulated points.  Where no blue curve is
visible, it is coincident with the red curve.
Experimental data (black and green circles) are compiled from:
$\gamma + p \to  p + \pi^0$: \cite{UKA97}--\cite{FUC96};
$\gamma + p \to  n + \pi^+$: \cite{UKA97}--\cite{GEN73},
                    \cite{ECK67}--\cite{MAC96};
$\gamma + d \to  n + p$: \cite{RPP04,ALE87}, \cite{COMPAS}--\cite{MOR89};
$\gamma + p \to  p + \pi^+ + \pi^-$: \cite{ALE87,CRO67}, 
                     \cite{BRA95}--\cite{ABB68};
$\gamma + p \to  p + 2 \pi^0$: \cite{BRA95}, \cite{KOT04}--\cite{HAR97};
$\gamma + p \to  n + \pi^+ + \pi^0$: \cite{BRA95,AHR03,LAN01};
$\gamma + p \to \Delta^{++} + \pi^-$: \cite{ALE87};
$\gamma + p$ total cross sections:  \cite{RPP04,ALE87,MAC96,COMPAS},
respectively. The green circles show recent
experimental data that became available to us after
we completed our fit; Although these recent
data agree reasonably well with our
approximations, a refitting would slightly improve the agreement.
}
\end{center}
\end{figure}

\noindent{\it New Approximations for Differential $\gamma + p$ Cross Sections}

The CEM2k \cite{CEM2k} version of CEM is a ``new generation" of the CEM 
following CEM97 \cite{CEM97}. Its development was partially motivated by 
the availability of some new, very precise and useful
experimental data obtained recently at GSI in Darmstadt, Germany, 
where a large number of measurements 
have been performed using inverse kinematics
for interactions of $^{56}$Fe,  $^{197}$Au, $^{208}$Pb, and $^{238}$U
at 1500, 1000, 800, 750, 500, and 300  MeV/nucleon with 
liquid $^1$H. These measurements provide
a large set of cross sections for production of practically
all possible isotopes from such reactions in a ``pure" form,
{\it i.e.}, individual cross sections from a specific given bombarding isotope
(or target isotope, when considering reactions in the usual kinematics,
p + A). Such cross sections are much easier to compare to models than the 
``camouflaged" 
data from $\gamma$-spectrometry measurements.
In addition, many reactions where a beam of light, medium,
or heavy ions with energy near to or below 1 GeV/nucleon interact with 
different nuclei, from the lightest, d, to the heaviest, $^{208}$Pb
were measured recently at GSI. References on these measurements and
many 
tabulated experimental cross sections may be found on the Web
page of Prof.\ K.-H. Schmidt \cite{SchmidtWebPage}. 

(We have analyzed with CEM2k and LAQGSM 
all measurements done at GSI of which we are aware, both for
proton-nucleus and nucleus-nucleus interactions; some examples of
our results compared with the GSI data and calculations by 
other available models may be found in
\cite{GSI03} and references therein.)

During the development of the CEM2k version of CEM and of LAQGSM, we
concentrated mainly on proton-nucleus and nucleus-nucleus reactions
and tried to improve the general description of different types of
nuclear reactions by our models, without focussing specifically on
photonuclear reactions. The main difference of CEM2k from its precursor 
CEM97 is in the criterion for when to move from the intranuclear-cascade
stage of a reaction to its preequilibrium stage, and when to move
from the latter to the evaporation/fission slow stage of the
reaction. In short, CEM2k has a longer cascade stage,
less preequilibrium emission, and a longer evaporation stage
with a higher excitation energy, as compared to CEM97 and CEM95.
Besides these changes to CEM97, we also made in CEM2k a 
number of other improvements and refinements, such as 
imposing momentum-energy conservation for each simulated event
(the Monte Carlo algorithm previously used in CEM 
provides momentum-energy conservation only 
statistically, on the average, but not exactly for each simulated
event); using real binding energies for nucleons at the cascade 
stage of a reaction instead of the approximation of a constant
separation energy of 7 MeV used in previous versions of the CEM; 
using reduced masses of particles in the calculation of their
emission widths instead of using the approximation
of no recoil used in previous versions; and
coalescence of complex particles from fast cascade nucleons
already outside the nucleus. On the whole, CEM2k describes better
than CEM97 and CEM95 many nuclear reactions, including the ones
induced by photons. CEM2k was incorporated by Gallmeier
into MCNPX to replace CEM97, and this version of MCNPX 
was extended by him to describe photonuclear reactions also in the
GDR region \cite{Franz04} (as a stand-alone code, CEM2k was developed
to describe photonuclear reactions only at energies
above the GDR region).

We have focused on the improved description of specifically the
photonuclear reactions when developing our latest version of CEM,
CEM03 \cite{OurNewINC}. Although CEM97 contained new approximations 
and agorithms to better describe
the integrated elementary $NN$, $\pi N$, and $\gamma N$ cross sections
as mentioned above, the double differential distributions of
secondary particles from such interactions were simulated by CEM2k
and all its precursors 
using the old Dubna INC approximations (7)--(11) for $\gamma p$
reactions (and similar relations, for $NN$ and $\pi N$ collisions).
These were obtained by Gudima {\it et al.} \cite{Dubna69,INC68a,INC68b} 
36 years ago, using the measurements available at that time.
In CEM03, for $NN$  and $\pi N$ collisions,
we addresed this problem by developing new approximations
similar to (7)--(11) and by using recent systematics by other authors,
based on experimental data available today (see details on $NN$
and $\pi N$ reactions in \cite{OurNewINC}). 
In the case of $\gamma p$ reactions (2) and (3), 
we chose another way:
Instead of fitting
the parameters $a_n$ from Eq.\ (7) at different 
$E_\gamma$ we found data
(see, {\it e.g.}, Figs.\ 2 and 3) and finding the energy dependence
of parameters $a_{nk}$ in Eq.\ (8) using the values obtained for $a_n$,
we took advantage of the event generator for $\gamma p$ and
$\gamma n$ reactions from the Moscow INC \cite{Iljinov97}
kindly sent us by Dr. Igor Pshenichnov. 
That event generator
includes a data file with smooth appproximations through presently
available experimental data 
at 50 different gamma energies from 117.65 to 6054 MeV (in the
system where the $p$ or $n$ interacting with $\gamma$ is at rest)
for the c.m.\ angular distributions 
$d \sigma / d \Omega$ of secondary particles as functions
of $\Theta$ tabulated for values of  $\Theta$ from 0 to 180 deg.,
with the step $\Delta \Theta = 10$ deg.,
for 60 different channels of $\gamma p$ and $\gamma n$ reactions 
considered by the Moscow INC (see details in \cite{Iljinov97}).
We use part of that data file with data for reactions (2) and (3),
and have written an algorithm to simulate unambigously
$d \sigma / d \Omega$ and to
choose the corresponding value of $\Theta$ for any $E_\gamma$,
using a single random number $\xi$ uniformly
distributed in the interval [0,1]. This is straightforward due to the 
fact that the function $\xi (\cos \Theta)$
$$
\xi(\cos \Theta) =
\int\limits_{-1}^{\cos \Theta}  d \sigma / d \Omega \: d \cos \Theta \bigg/\!\!
\int\limits_{-1}^{1}  d \sigma / d \Omega \: d \cos \Theta $$
is a smooth monotonic function increasing from 0 to 1  as
$\cos \Theta$ varies from -1 to 1. Naturally, when $E_\gamma$
differs from the values tabulated in the data file, we perform
first the needed interpolation in energy.
We use this procedure
to describe in CEM03 angular distributions of secondary particles from
reactions (2) and (3), as well as for isotopically symmetric
reactions $\gamma + n \to n + \pi^0$ and $\gamma + n \to p + \pi^-$.

Examples of eight angular distributions of $\pi^{0}$ from
$\gamma p \to \pi^{0} p$  and 
of $\pi^{+}$ from $\gamma p \to \pi^{+} n$ 
as functions of $\Theta^{\pi}_{c.m.s}$ are
presented in Figs.\ 2 and 3. We see that the approximations
developed in CEM03 (solid histograms)
agree much better with the available experimental data
than the old Dubna INC approximations (7)--(8)
used in all precusors of CEM03 (dashed histograms).

\noindent{\bf New Approximations for $\gamma + A$ Absorptrion Cross Sections}

CEM03 (and its predecessors) does not consider 
absorption of low energy photons in the GDR region and
takes into account photoproduction on free nucleons of only two pions.
This restricts its applicability to the range
30 MeV$ \lesssim E_\gamma \lesssim 1.5$ GeV, which is not convenient 
when it is used as an event generator in a transport code.

To extend the applicability of CEM03 (and LAQGSM03)
into the GDR region, it is necessary
to omit the intranuclear cascade (INC) and to
consider such reactions as starting with the preequilibrium model.
The INC used by CEM03 as the first stage of arbitrary reactions
is a semiclassical model that does not consider any
collective degrees of freedom of a nucleus, including the GDR; in addition,
the energy of a $\gamma$ in the GDR region is too low 
to justify the use of any INC. In our approach, it is possible 
to deal with this limitation as was done 
30 years ago \cite{Lukyanov75}, using the
Modified Exciton Model (MEM) \cite{MEM,MODEX}
used in the initial version of CEM \cite{CEM} 
and 25 years later \cite{Belov00}, using an improved version 
of the MEM contained in the CEM95 \cite{CEM95} code.
We plan to extend CEM03 to describe photonuclear
reactions in the GDR region in the near future, but this requires
a large amount of tedious work: 1) to make sure that we use
the most reliable parameters of the GDR for all nuclei, 2) to define
an optimal transition from the current three-stage (INC, preequilibrium,
and evaporation/fission) description of reactions
to a two-stage approach needed in the GDR region, and 3) to test the
extended model against available experimental data.

The describe properly with CEM03 and LAQGSM03 photonuclear reactions above 
$E_\gamma \sim 1.5$ GeV,
it is necessary to take into account production of more tham two pions
in $\gamma N$ colissions, as well as to consider production of resonances
heavier than $\Delta(1232)$, as has been done, {\it i.e.}, in the
Moscow INC \cite{Iljinov97}. We plan to extend CEM03 and LAQGSM03
to higher $E_\gamma$ in subsequent versions.

\begin{figure}[!ht]\begin{center}
\includegraphics[width=170mm]{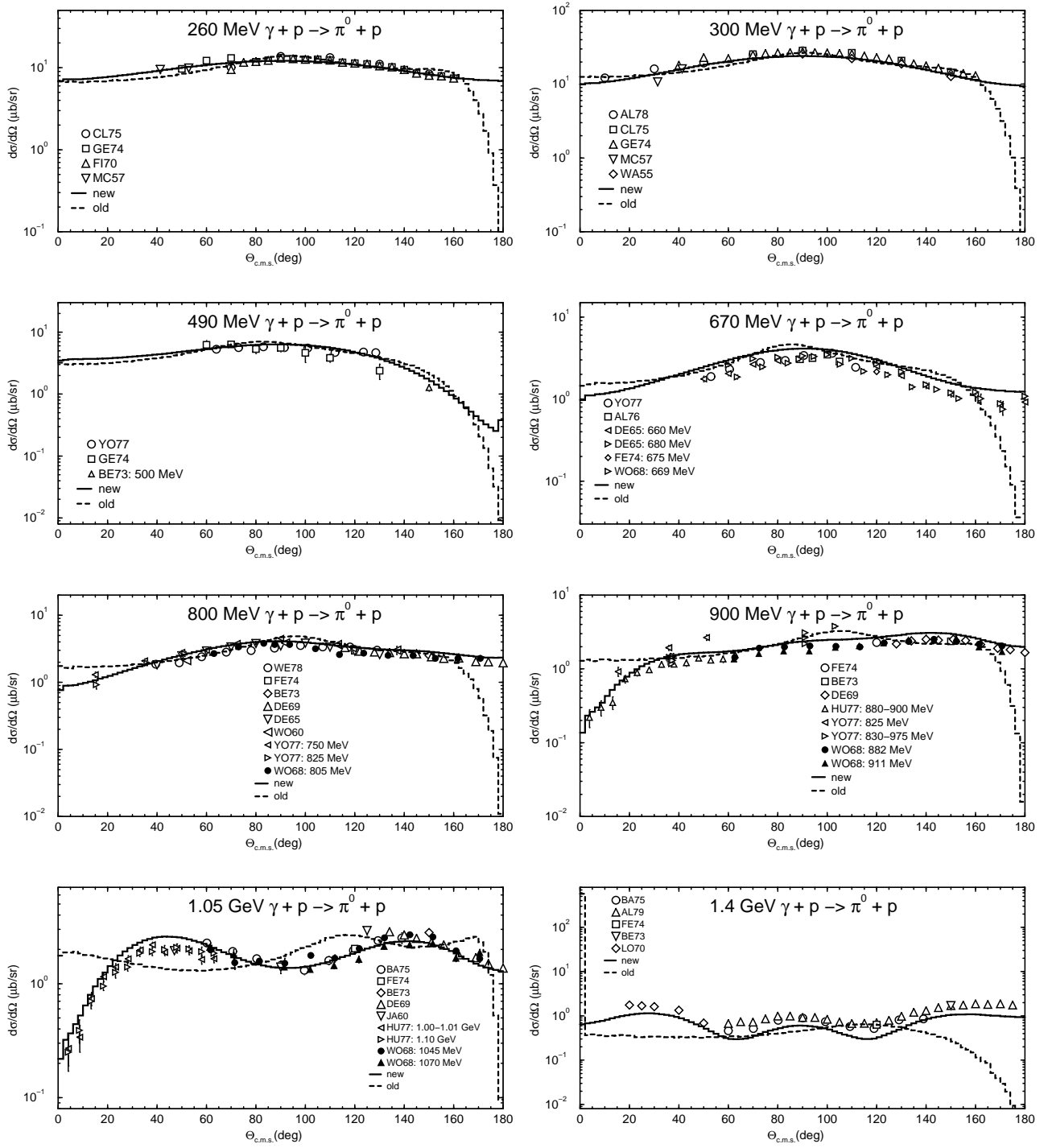}
\caption{
Example of eight
angular distributions of $\pi^{0}$ from
$\gamma p \to \pi^{0} p$  
as functions
of $\Theta^{\pi}_{c.m.s}$ at photon energies from 260 MeV to 1.4 GeV.
The dashed lines show the old approximations used in the Dubna INC PRM 
while the solid lines are
our new approximations incorporated into the 
CEM03 and LAQGSM03 codes. Experimental data are shown by symbols and are
from:
CL75 \cite{CL75},
GE74 \cite{GE74},
FI70 \cite{FI70},
MC57 \cite{MC57},
AL78 \cite{AL78},
WA55 \cite{WA55},
YO77 \cite{YO77},
BE73 \cite{BE73},
AL76 \cite{AL76},
DE65 \cite{DE65},
FE74 \cite{FE74},
WO68 \cite{WO68},
WE78 \cite{WE78},
DE69 \cite{DE69},
WO60 \cite{WO60},
HU77 \cite{HU77},
BA75 \cite{BA75},
JA60 \cite{JA60},
AL79 \cite{AL79}, and
LO70 \cite{LO70}; 
tabulated values are available at: 
http://www-spires.dur.ac.uk/hepdata/reac2.html.
}
\end{center}
\end{figure}

\begin{figure}[!ht]\begin{center}
\includegraphics[width=170mm]{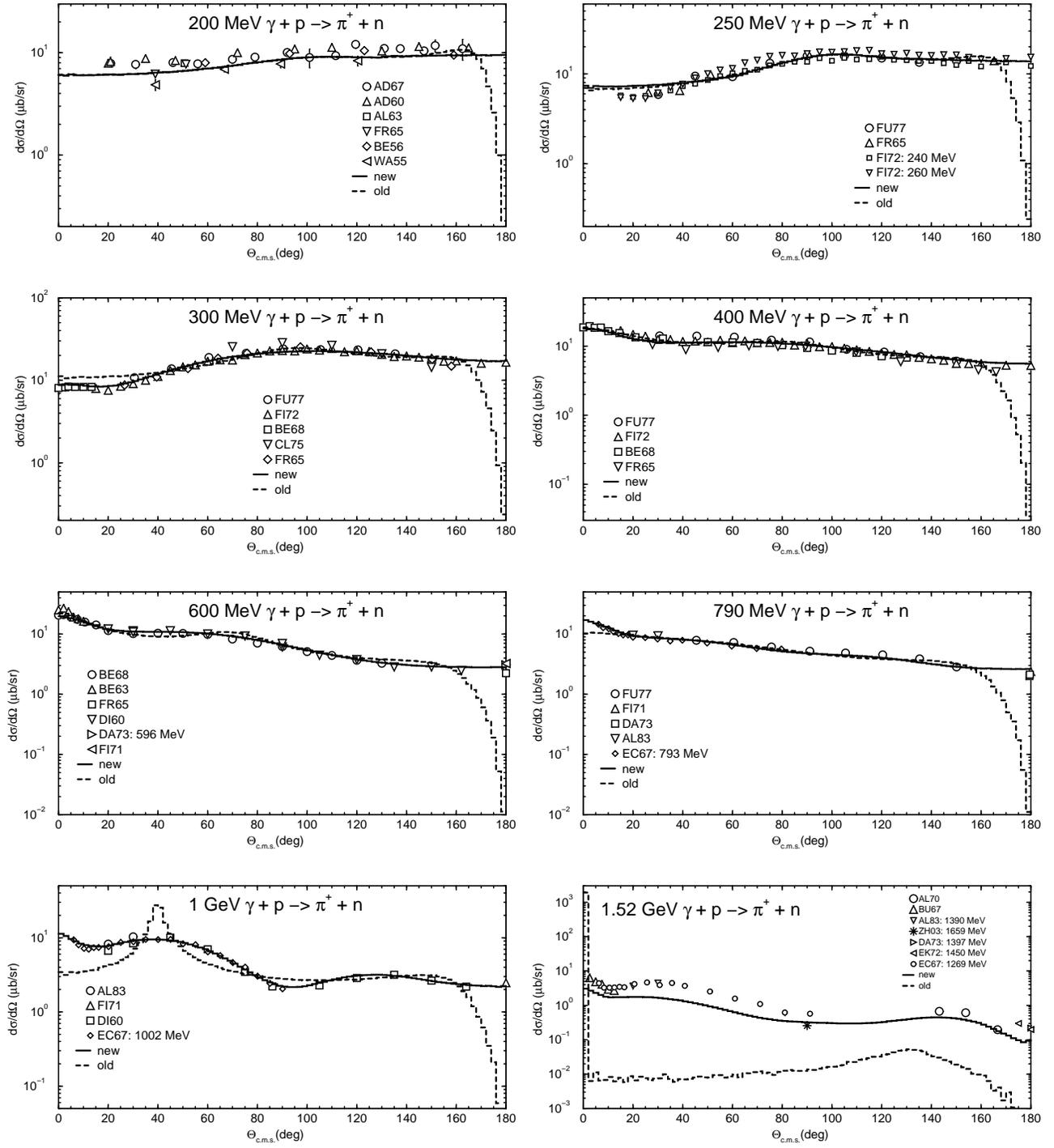}
\caption{
Example of eight
angular distributions of $\pi^{+}$ from
$\gamma p \to \pi^{+} n$ 
as functions
of $\Theta^{\pi}_{c.m.s}$ at photon energies from 200 MeV to 1.52 GeV.
The dashed lines show the old approximations used in the Dubna INC PRM 
while the solid lines are
our new approximations incorporated into the 
CEM03 and LAQGSM03 codes. Experimental data are shown by symbols and are
from:
AD67 \cite{AD67},
AS60 \cite{AD60},
AL63 \cite{AL63},
FR65 \cite{FR65},
BE56 \cite{BE56},
WA55 \cite{WA55a},
FU77 \cite{FU77},
FI72 \cite{FI72},
BE68 \cite{Be68},
CL75 \cite{CL75}, 
BE63 \cite{BE63},
DI60 \cite{DI60},
DA73 \cite{DA73},
FI71 \cite{FI71},
AL83 \cite{AL83},
EC67 \cite{EC67},
AL70 \cite{AL70},
BU67 \cite{BU67},
ZH03 \cite{ZH03}, and
EK72 \cite{EK72};
tabulated values are available at: 
http://www-spires.dur.ac.uk/hepdata/reac2.html.
}
\end{center}
\end{figure}

In the meantime, for applications it is possible to
get quite reasonable results for spectra of emitted nucleons 
and complex particles and for the nuclide production cross sections 
with our present CEM03 model for photonuclear reactions both in 
the GDR region and at $E_\gamma \gtrsim 1.5$ GeV, by employing a
correct total photoabsorption cross section. Indeed, CEM03 
starts a reaction in the GDR region with a cascade and since
the $\gamma$ energy is below the pion-production threshold,
the only available reaction channel is to absorb such photons 
on a quasideuteron pair of nucleons, generating two ``cascade" 
nucleons inside the nucleus.
As the energy of these nucleons is low, $\sim 10$ MeV, 
these nucleons are ``absorbed" by the nucleus generating
two excited nucleons (excitons) and two holes, then CEM03 would
proceed with this process as a preequilibrium reaction
followed by evaporation/fission. All the real calculation
of such reactions would be done
with only the preequilibrium and evaporation models and the INC would
serve only to provide the number of excitons, as an input to the MEM. 
At the end of the calculation, the total photoabsorption
(reaction) cross section is needed to normalize the results. 
CEM03 (and most other INC models) calculates the 
total reaction cross section, $\sigma_{in}$, by the Monte Carlo method
using the geometrical cross section, $\sigma_{geom}$, and the
number of inelastic, $N_{in}$, and elastic, $N_{el}$, simulated
events, namely: 
$\sigma_{in} = \sigma_{geom} N_{in} / (N_{in} + N_{el})$.
This approach provides a good agreement with available data 
for reactions induced by nucleons, pions, and photons at incident
energies above about 100 MeV, but is not reliable at 
energies below 100 MeV
(see, {\it e.g.}, Figs.\ 3 and 4 and Ref.\ \cite{Mashnik02c}). 

To address this problem for photonuclear reactions, 
we have written a FORTRAN routine GABS based on
the recent approximation by Kossov \cite{Kossov}, that
provides reliable photoabsorption cross sections on
arbitrary targets at all energies from the hadron production
threshold to about 40 TeV. We have added GABS to CEM03
to normalize our photonuclear results 
to this systematics rather than to  $\sigma_{in}$
calculated by the Monte Carlo method,
as we have done previously.
(As a rule, we use LAQGSM03 only at energies above several GeV,
where CEM03 becomes already not reliable; at such high energies, 
LAQGSM03 describes quite well $\sigma_{in}$ and does not require
renormalization of its results to any systematics;
therefore we do not incorporate GABS into LAQGSM03.)

The Kossov approximation \cite{Kossov} of the energy dependence of
photonuclear cross sections is subdivided into three main regions:
the GDR region, the nucleon resonance region, and the high-energy region.
Its functional form is also subdivided into three groups depending on
the mass number of the target: the  $\sigma_{\gamma p}$ cross section,
the cross section for  $\gamma d$ reactions, and the  $\sigma_{\gamma A}$ 
cross section for $A > 2$.

The Kossov approximation \cite{Kossov}  
of $\sigma_{\gamma p}$ (in mb) as a 
function of the photon energy $E$ (in MeV) 
is of the following form:
\begin{equation}                                                   
\sigma_{\gamma p}=f_r \cdot (r_\Delta + r_H) + g_4 + g_8 + f_p \cdot h_p^{(p)}
\mbox{ ,}
\end{equation}
where
\begin{equation}                                                   
f_r = (1 + e^{25 \cdot (5.24 - z )} )^{-1} \mbox{ ,}
\end{equation}
\begin{equation}                                                   
r_\Delta = {0.55 \over{1 + {{(z-u_\Delta(1))^2}\over{w_\Delta(1)} }}} \mbox{ ,} 
\end{equation}
\begin{equation}                                                   
u_\Delta(A) = 5.82 - {0.07 \over{1 + 0.003 \cdot A^2}} \mbox{ ,} 
\end{equation}
\begin{equation}                                                   
w_\Delta (A) = 0.056 + \ln (A) \cdot ( 0.03 - 0.001 \cdot \ln (A)) \mbox{ ,}
\end{equation}
\begin{equation}                                                   
r_H = {0.223\over{1 + {(z-6.57)^2\over{w_H(1)}} } } \mbox{ ,}
\end{equation}
\begin{equation}                                                   
w_H(A) = 0.045 + 0.04 \cdot (\ln(A))^{3\over2} \mbox{ ,}
\end{equation}
\begin{equation}                                                   
g_4 = {e^{4\cdot (6.27-z)}\over{1 + e^{12 \cdot (7.25 -z)}} }  \mbox{ ,}
\end{equation}
\begin{equation}                                                   
g_8 = {e^{8\cdot (6.66-z)}\over{1 + e^{24 \cdot (6.9  -z)}} }  \mbox{ ,}
\end{equation}
\begin{equation}                                                   
f_p = (1 + e^{4\cdot (7-z)})^{-1} \mbox{ ,}
\end{equation}
\begin{equation}                                                   
h_p^{(p)} = 0.0375 \cdot (z - 16.5) + 1.07 \cdot e^{-0.11 \cdot z} \mbox{ ,}
\end{equation}
and $z = \ln(E)$.

For $\gamma d$ reactions, the Kossov approximation is as follows:
\begin{equation}                                                   
\sigma_{\gamma d}=f_r \cdot (r_\Delta + r_H) + g_1 + g_2 + g_4 + g_8 
+ s_p(2) f_p h_p(2) 
\mbox{ ,}
\end{equation}
where
\begin{equation}                                                   
f_r = (1 + e^{25 \cdot (\tau_r(2) - z )} )^{-1} \mbox{ ,}
\end{equation}
\begin{equation}                                                   
\tau_r(A) = 5.13 - 0.00075 \cdot A \mbox{ ,}
\end{equation}
\begin{equation}                                                   
r_\Delta= {0.88 \over{1 + {{(z-u_\Delta(2))^2}\over{w_\Delta(2)} }}}\mbox{ ,} 
\end{equation}
\begin{equation}                                                   
r_H = {0.348\over{1 + {(z-6.575)^2\over{w_H(2)}} } } \mbox{ ,}
\end{equation}
\begin{equation}                                                   
g_1 = {e^{1\cdot (1.86-z)}\over{1 + e^{3 \cdot (1.2 -z)}} }  \mbox{ ,}
\end{equation}
\begin{equation}                                                   
g_2 = {e^{2\cdot (2.11-z)}\over{1 + e^{6 \cdot (1.5 -z)}} }  \mbox{ ,}
\end{equation}
\begin{equation}                                                   
g_4 = {e^{4\cdot (6.2 -z)}\over{1 + e^{12\cdot (7.1 -z)}} }  \mbox{ ,}
\end{equation}
\begin{equation}                                                   
g_8 = {e^{8\cdot (6.62-z)}\over{1 + e^{24\cdot (6.91-z)}} }  \mbox{ ,}
\end{equation}
\begin{equation}                                                   
s_p(A) = A \cdot ( 1 - 0.072 \cdot \ln(A)) \mbox{ ,}
\end{equation}
\begin{equation}                                                   
h_p(A) = 0.0375 \cdot (z - 16.5) + s_h(A) \cdot e^{-0.11 \cdot z} \mbox{ ,}
\end{equation}
\begin{equation}                                                   
s_h(A) = 1.0663  - 0.0023\cdot \ln(A) \mbox{ .}
\end{equation}

For $\gamma A$ reactions, when $A > 2$, the Kossov approximation is 
similar to Eq.\ (23) but has a different functional form, therefore it
is more convenient to write it as follows:
\begin{equation}                                                   
\sigma_{\gamma A} = \sigma_{GDR} + f_r(r_\Delta + r_H) + s_p(A) f_p h_p(A)
\mbox{ ,} 
\end{equation}
where the ``global" approximation for the photoabsorption cross
section in the GDR region can be writen as
\begin{equation}                                                   
\sigma_{GDR} = g_1 + g_2 + g_4 + g_8 \mbox{ ,}
\end{equation}
\begin{equation}                                                   
g_i = {e^{i \cdot (\rho_i - z)}\over{1 + e^{3 i \cdot (\tau_i -z)}}} \mbox{ ,}
\end{equation}
\begin{equation}                                                   
\rho_1 = { {3.2+0.75 \cdot \ln(A)}\over{1 + (2/A)^4}} \mbox{ ,}
\end{equation}
\begin{equation}                                                   
\tau_1 = { {6.6-0.5  \cdot \ln(A)}\over{1 + (2/A)^4}} \mbox{ ,}
\end{equation}
\begin{equation}                                                   
\rho_2 = { {4.0+0.125\cdot \ln(A)}\over{1 + (2/A)^4}} \mbox{ ,}
\end{equation}
\begin{equation}                                                   
\tau_2 = { 3.4\over{1 + (2/A)^4}} \mbox{ ,}
\end{equation}
\begin{equation}                                                   
\rho_4 = 3.8 + 0.05 \cdot \ln(A) \mbox{ ,}
\end{equation}
\begin{equation}                                                   
\tau_4 = 3.8 - 0.25 \cdot \ln(A) \mbox{ ,}
\end{equation}
\begin{equation}                                                   
\rho_8 = 3.65- 0.05 \cdot \ln(A) \mbox{ ,}
\end{equation}
\begin{equation}                                                   
\tau_8 = 3.5 - 0.16 \cdot \ln(A) \mbox{ .}
\end{equation}

We note that Eq.\ (45) was misprinted in the original publication by 
Kossov \cite{Kossov} (it corresponds to
Eq.\ (41) in \cite{Kossov}) where a ``$+$" sign occurs
instead of a ``$-$" sign. The misprinted formula does
not reproduce the cross sections presented in \cite{Kossov}, whereas
the corrected version does. 

Figs.\ 4 and 5 show examples of twelve photoabsorption cross sections
on several light, medium, and heavy nuclei. In these
figures,  we compare predictions of the Kossov systematics as
implemented in the routine GABS with available experimental data
and with the LANL, KAERI, and the BOFOD(MOD) (IPPE/Obninsk and CDFE/Moscow) 
evaluations from the IAEA Photonuclear Data Libtary \cite{IAEA}, as well as
with calculations by two older versions of CEM, namely, the CEM95 
photonuclear code version \cite{CEM95photo} and 
CEM2k as modified by Gallmeier \cite{Franz04} for MCNPX.

The Kossov systematics describe well the experimental
photoabsorption cross sections and agree with the LANL, KAERI,
and the BOFOD evaluations, especially for heavy targets. For $^{12}$C,
$^{27}$Al, and $^{63}$Cu 
(and several other nuclei we tested but did not
include in Figs.\ 4 and 5) the agreement in the GDR region is not so
good. This is because we use here the ``global" approximation given
by Eqs.\ (36)--(45) to calculate the photoabsorption cross section in the
GDR region for all nuclei. It is known from the literature that
the GDR of light nuclei differ significantly from the ones of heavy
nuclei, and should be addresed carefully for each light nucleus
separately. In fact, Kossov  \cite{Kossov}
had fitted the parameters of the light nuclei separately
and his results shown in Figs.\ 2--7 of Ref.\ \cite{Kossov} 
for the light nuclei agree better with the data than the
``global" systematics shown here does. Unfortunately,  Kossov  
did not publish in  \cite{Kossov} the parametrization of the GDR 
he found for every light nucleus (some details of this are listed
in the recent {\it GEANT4 Physics Reference Manual} \cite{GEANT4_04} 
and in \cite{Wellisch03},
but only for some light nuclei, and those details differ from what 
is published in \cite{Kossov}). To fill this gap,
we hope to determine ouselves a parameterization of the GDR 
photoabsorption on light nuclei using all available 
experimental data. 

\noindent{\bf Illustrative Results}

In this Section, we present several illustrative results from
CEM03 and LAQGSM03 extended to describe photonuclear
reactions. We start with photofission cross sections, which
we have compiled from the literature and have analyzed with both
CEM03 and LAQGSM03. Fig.\ 6 shows a comparison of such data
for $^{197}$Au, $^{208}$Pb, $^{209}$Bi, $^{232}$Th,
$^{233}$U, $^{235}$U, $^{238}$U, and $^{237}$Np to results
of CEM03, as well as to several earlier versions, namely,
the photonuclear versions of CEM95 \cite{CEM95photo},
CEM98 \cite{CEM98}, CEM2k+GEM2 \cite{fitaf},
and the modified version of CEM2k incorporated into MCNPX by 
Gallmeier \cite{Franz04}. 
The CEM03 results agree well with the experimental 
fission cross sections, and better than the results of the earlier
models.
Using in CEM03 (and CEM2k+GEM2) the Kossov approximation for the total
photoabsorption cross sections allows us to describe the fission
cross section not only for photon energies from
$\sim 30$ MeV to $\sim 1.5$ GeV, where the model is expected to be
reliable, but also outside this region, in the whole
range from 10 MeV to 5 GeV. 

\newpage
\begin{figure}[!h]
\begin{center}
\includegraphics[width=15.5cm]{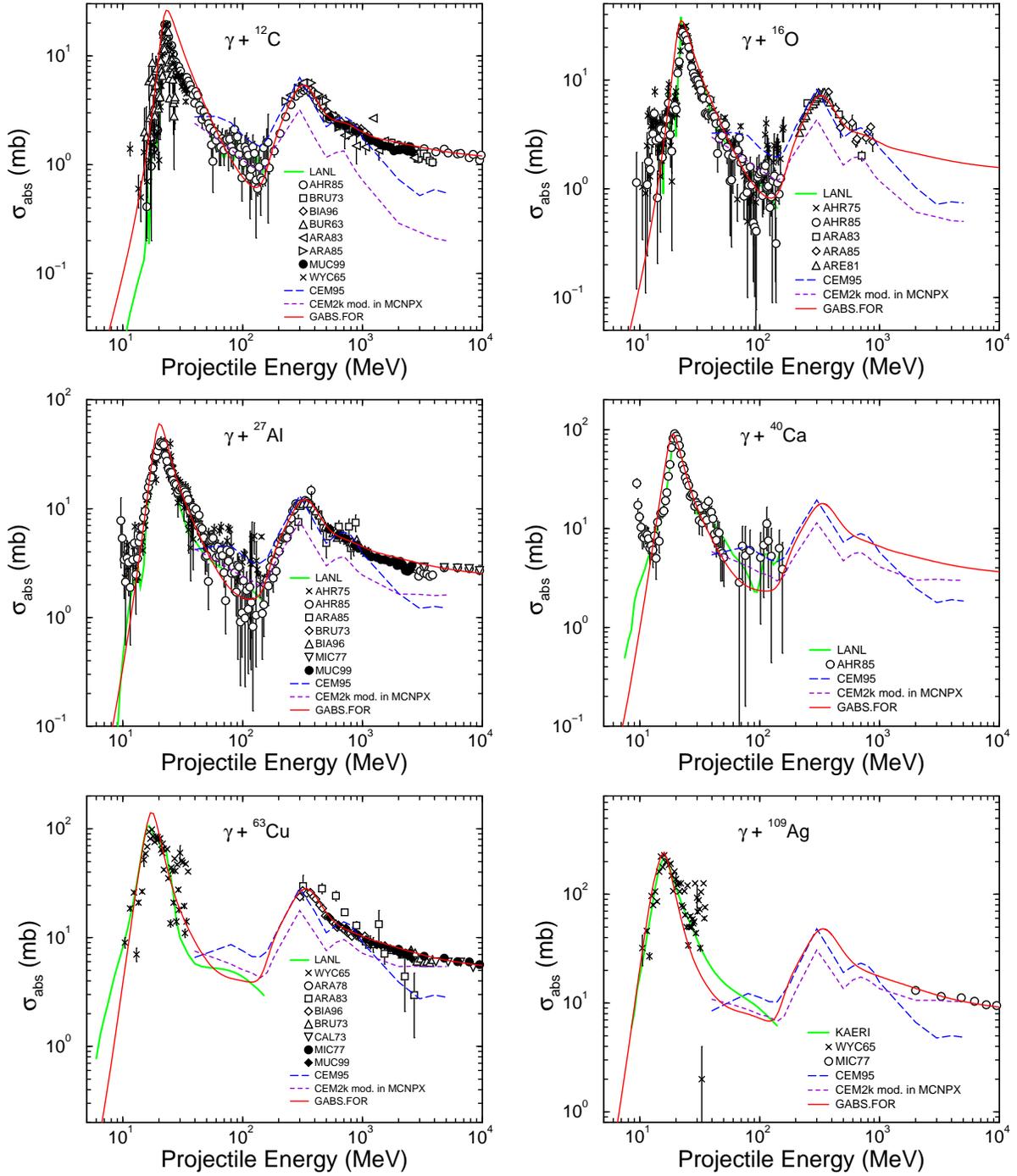}
\caption{
Examples of total photoabsorption cross sections for 
$^{12}$C, $^{16}$O, $^{27}$Al, $^{40}$Ca, $^{63}$Cu, and $^{109}$Ag
as functions of photon energy. The red lines marked as ``GABS.FOR" are
from our subroutine written to reproduce Kossov's \cite{Kossov}
systematics, as described in the text. 
The green line marked as ``LANL" (or ``KAERI", for $^{109}$Ag) show the
evaluations by LANL 
(or KAERI, for $^{109}$Ag) 
from the IAEA Photonuclear Data Library
\cite{IAEA}. Results from the 
photonuclear version of CEM95 \cite{CEM95photo} and
from CEM2k as modified for MCNPX by Gallmeier \cite{Franz04}
are shown by the blue and brown dashed lines, respectively.
Experimental data (symbols) are from: 
AHR85 \cite{AHR85},
BRU73 \cite{BRU73},
BIA96 \cite{BIA96},
BUR63 \cite{BUR63},
ARA83 \cite{ARA83},
ARA85 \cite{ARA85},
MUC99 \cite{MUC99},
WYC65 \cite{WYC65},
AHR75 \cite{AHR75},
ARE81 \cite{ARE81},
MIC77 \cite{MIC77},
ARA78 \cite{ARA78}, and
CAL73 \cite{CAL73}.
}
\end{center}
\end{figure}

\newpage

\begin{figure}[!h]
\begin{center}
\vspace*{-5mm}
\includegraphics[width=15.0cm]{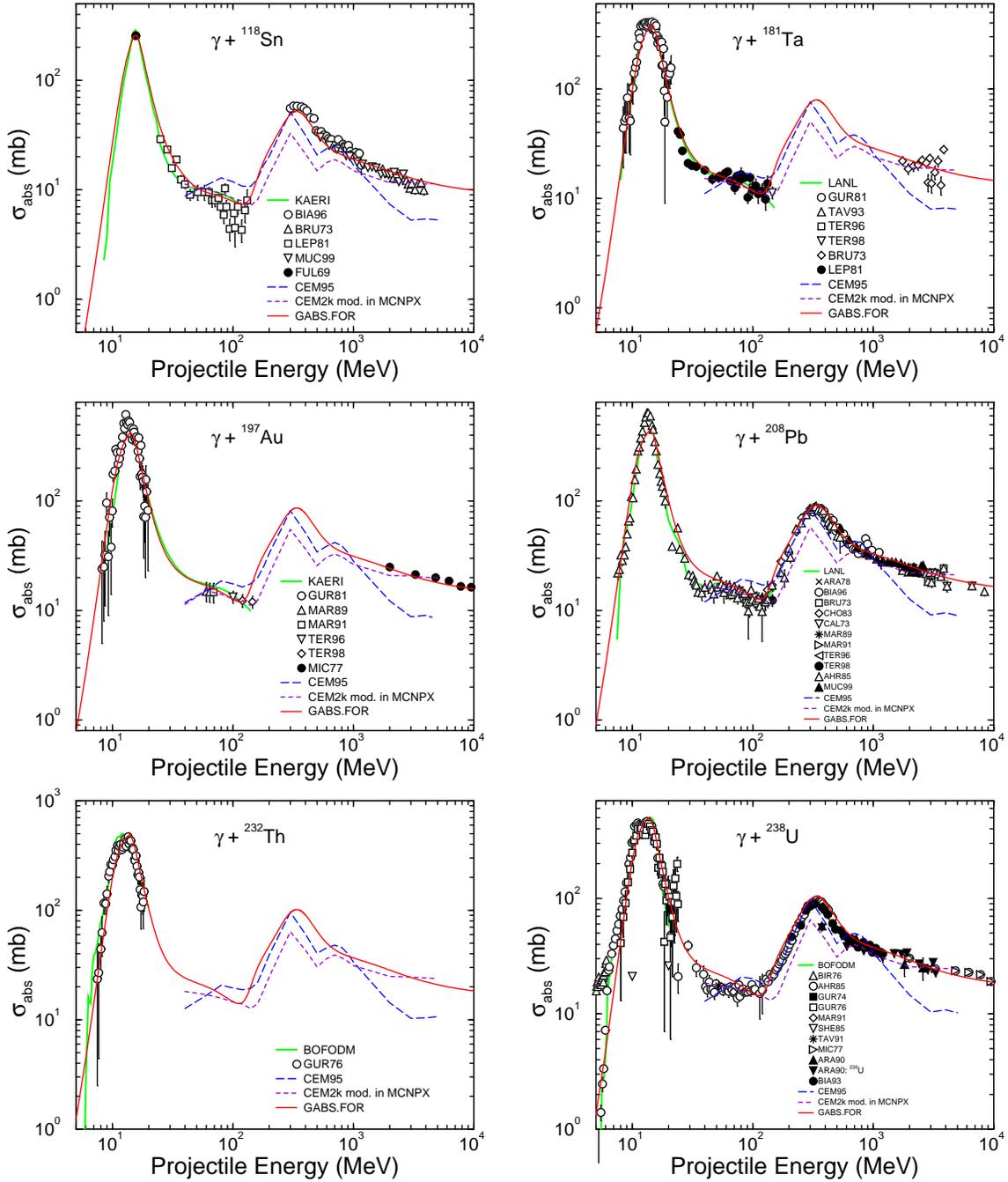}
\caption{
Examples of total photoabsorption cross sections for 
$^{118}$Sn, $^{181}$Ta, $^{197}$Au, $^{208}$Pb, $^{232}$Th, and $^{238}$U
as functions of photon energy. The red lines marked as ``GABS.FOR" are
results by our subroutine written to reproduce Kossov's \cite{Kossov}
systematics, as described in the text. 
The green line marked as ``LANL", ``KAERI", or ``BOFODM" 
show the
evaluations by LANL, KAERI, or by a collaboration between IPPE/Obninsk
and CDFE/Moscow (the BOFOD(MOD) Library) from the IAEA Photonuclear Data 
Library \cite{IAEA}. Results from the 
photonuclear version of CEM95 \cite{CEM95photo} and
from CEM2k as modified for MCNPX by Gallmeier \cite{Franz04}
are shown by the blue and brown dashed lines, respectively.
Experimental data (symbols) are from: 
BIA96 \cite{BIA96}, 
BRU73 \cite{BRU73}, 
LEP81 \cite{LEP81},
MUC99 \cite{MUC99}, 
FUL69 \cite{FUL69},
GUR81 \cite{GUR81},
TAV93 \cite{TAV93},
TER96 \cite{TER96}, 
TER98 \cite{TER98}, 
MAR89 \cite{MAR89}, 
MAR91 \cite{MAR91}, 
MIC77 \cite{MIC77}, 
ARA78 \cite{ARA78}, 
CHO83 \cite{CHO83},
CAL73 \cite{CAL73}, 
AHR85 \cite{AHR85}, 
GUR76 \cite{GUR76},
BIR76 \cite{BIR76},
GUR74 \cite{GUR74},
SHE85 \cite{SHE85},
TAV91 \cite{TAV91},
ARA90 \cite{ARA90}, and
BIA93 \cite{BIA93}.
}
\end{center}
\end{figure}

\newpage
\begin{figure}[!h]
\begin{center}
\includegraphics[width=16.5cm]{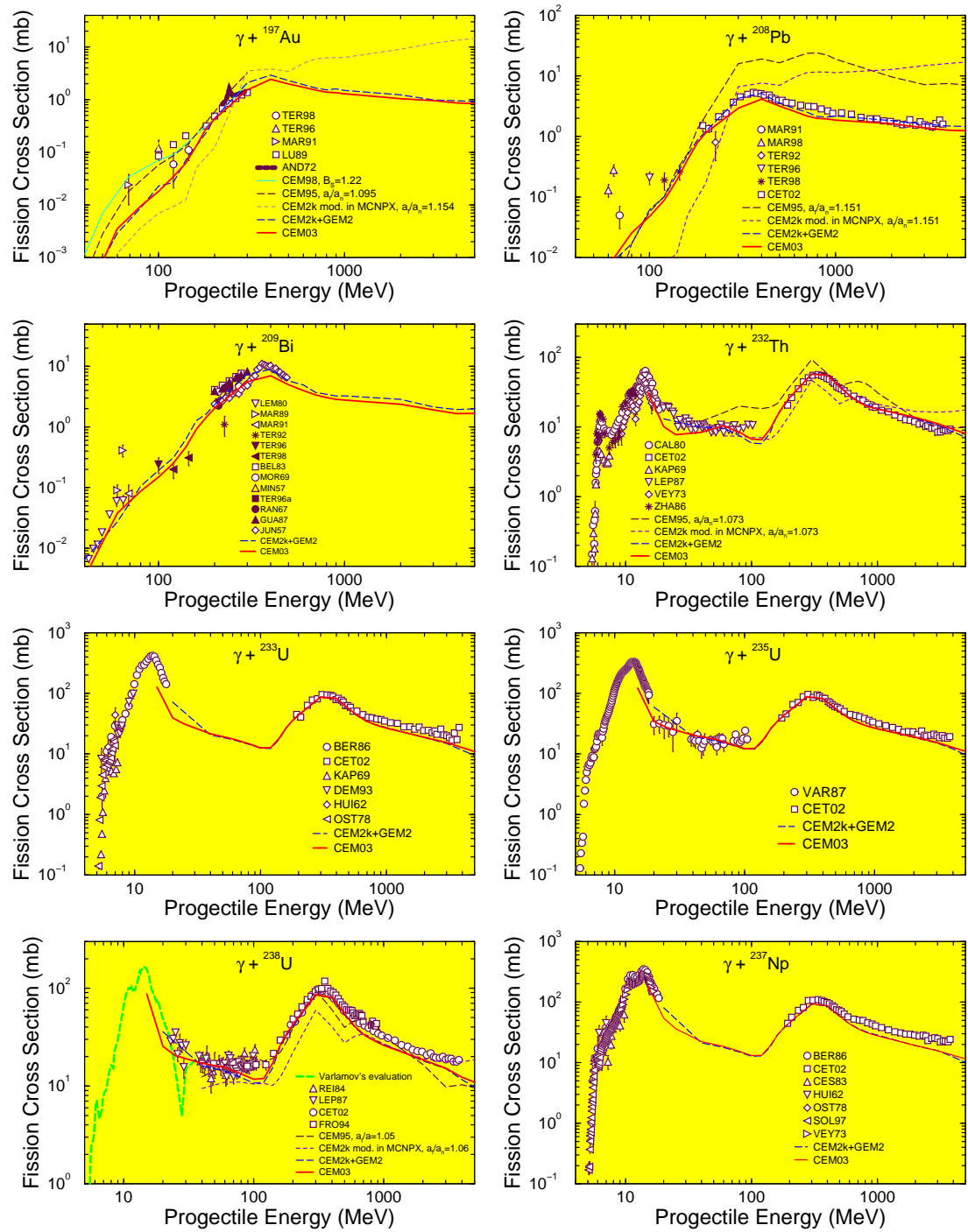}
\vspace{-30mm}
\caption{
Comparison of calculated photofission cross sections
on $^{197}$Au, $^{208}$Pb, $^{209}$Bi, $^{232}$Th,
$^{233}$U, $^{235}$U, $^{238}$U, and $^{237}$Np
with experimental data (symbols), 
results by previous versions of CEM (see details and references in
the text), and the statistical evaluation by Varlamov {\it et al.}
from independent measurements
\cite{VarEv},
as indicated. Experimental data are from: 
TER98 \cite{TER98}, 
TER96 \cite{TER96}, 
MAR89 \cite{MAR89}, 
MAR91 \cite{MAR91}, 
LU89  \cite{LU89},
AND72 \cite{AND72},
TER92 \cite{TER92},
CET02 \cite{CET02},
LEM80 \cite{LEM80},
BEL83 \cite{BEL83},
MOR69 \cite{MOR69},
MIN57 \cite{MIN57},
TER96a \cite{TER96a},
RAN67 \cite{RAN67},
GUA87 \cite{GUA87},
JUN57 \cite{JUN57},
CAL80 \cite{CAL80},
KAP69 \cite{KAP69},
LEP87 \cite{LEP87},
VEY73 \cite{VEY73},
ZHA86 \cite{ZHA86},
BER86 \cite{BER86},
DEM93 \cite{DEM93},
HUI62 \cite{HUI62},
OST78 \cite{OST78},
VAR87 \cite{VarEv},
REI84 \cite{REI84},
FRO94 \cite{FRO94},
CES83 \cite{CES83},
OST78 \cite{OST78}, and
SOL97 \cite{SOL97}.
}
\end{center}
\end{figure}

\newpage
This is a example of getting reasonably good
results outside the region where the model is justified, 
as discussed in the previous Section.
Results of LAQGSM03 for these fission cross sections practically
coincide with the ones by CEM03 above the GDR region, as
the calculation of fission cross sections in CEM03 and LAQGSM03
were developed to be (see details in \cite {fitaf}), but are
significantly lower than the data in the GDR region, as LAQGSM03
does not use the Kossov approximation and so should not be applied
in the GDR region. Results of CEM95, CEM98, and CEM2k
are also below the data in the GDR region, for the same reason.

We note that all the CEM03 and LAQGSM03
results shown in Fig.\ 6 and in the follwing figures are
obtained using default and fixed values of all parameters,
without fitting anything. We only specify in the inputs to CEM03
(and CEM2k+GEM2) and LAQGSM03 the energy of the incident 
photons and $A$ and $Z$ of the target, then calculate.
CEM95, CEM98, and CEM2k use a parameter whose value affects drastically
the calculated fission cross sections, just as in many similar
statistical models: This is the ratio of the level density parameters 
used in the
fission and evaporation channels, $a_f/a_n$ (or, $B_s$, in the case
of CEM98, see details in \cite{CEM98}). The fission cross sections 
calculated by any code employing the statistical evaporation and fission 
models depend so much on $a_f/a_n$ that by fitting this ratio
it is possible to get a good agreement with the measured data
(but not to predict unmeasured fission cross sections) with
any reasonable values for the fission barriers,
nuclide masses, pairing energies, and deformations, for any
particular measured reaction. This is why some published papers that
analyze fission cross sections or even pretend to obtain
``experimental fission barriers" 
without addresing the question of $a_f/a_n$
are of low significance. 
In our CEM95, CEM98, and CEM2k calculations,
we use the default options for nuclear masses, pairing energies, and  
fission barriers (the ``recommended" options, in the case of CEM95, where 
several options are available in its input; see details in \cite{CEM95}),
but we still need to define (more exactly, to fit) the values of
$a_f/a_n$ (or, $B_s$, in the case of CEM98): These values are listed
on our plots in Fig.\ 6. 
Naturally, CEM2k+GEM2, CEM03, and LAQGSM03 also had in the beginning
the problem
of $a_f/a_n$, but this problem was solved in \cite{fitaf}
by fitting these parameters for proton-induced fission cross sections 
for all targets for which we found data, at all incident energies,
and by their extrapolation/interpolation for unmeasured targets.
The fitted values are fixed and are used in all our 
further CEM03 and LAQGSM03 calculations without subsequent variation;
this allows us not only to describe well most of the measured data but 
also to predict reasonably well unmeasured fission cross sections.
The fitting procedure \cite{fitaf} 
was done so that both CEM03 (and CEM2k+GEM2) and LAQGSM03
describe as well as possible all available
proton-induced measured fission cross sections;
this is why the fission cross sections calculated by CEM03 practically
coincide with the ones obtained by LAQGSM03 and with the experimental data.

We note that both CEM03 and LAQGSM03 
assume that the reactions occur generallly in three stages
(see {\it e.g.} \cite{Mashnik04}). 
The first stage is the IntraNuclear Cascade (INC),
in which primary particles can be re-scattered and produce secondary
particles several times prior to absorption by, or escape from the nucleus.
When the cascade stage of a reaction is completed, both our codes use the
coalescence model described in \cite{Toneev83}
to ``create" high-energy d, t, $^3$He, and $^4$He by
final-state interactions among emitted cascade nucleons, already outside 
of the target.
 The emission of the cascade particles determines the particle-hole 
configuration, Z, A, and the excitation energy that is
the starting point for the second, pre-equilibrium stage of the
reaction.  
The subsequent relaxation of the nuclear excitation is
treated in terms of the modified exciton model of pre-equilibrium decay 
followed by the equilibrium evaporation/fission stage of the reaction.
Generally, all four components may contribute to experimentally measured 
particle spectra and distributions. 
But if the residual nuclei after the INC have atomic numbers 
with  $A \le 12$,  both CEM03 and LAQGSM03 use the Fermi break-up model 
described in \cite{LAQGSM} to calculate their further disintegration 
instead of using the preequilibrium and evaporation models. 
The Fermi break-up is much faster than, and gives results very similar 
to, the continuation of the more detailed models to
much lighter nuclei.

Figs.\ 7 and 8 show two examples of proton spectra calculated by
CEM03 and LAQGSM03 compared with experimental data for the reactions
300 MeV $\gamma$ + Cu \cite{SCH82} and
198 MeV $\gamma$ + C \cite{ANG86}, respectively. Both codes
describe quite well the proton spectra in the case of copper, 
but less well for carbon.

Fig.\ 9 shows expamles of $\pi^+$ angular distributions from 213 MeV
$\gamma$'s interacting with Pb, Sn, Ca, and C targets. One can see that 
the $\pi^+$ angular distributions calculated by CEM03
agree reasonably well with the experimental data \cite{Fissum96}
for C, Ca, and Sn targets, but
underestimate by a factor of 2 to 3 the Pb data. 
We do not have a good understanding of this disagreement. 
One possible explanation
of this would be if CEM03 absorbs too strongly the low-energy pions produced
in $\gamma p$ collisions inside the target. The heavier the target
the bigger would be this effect, thefore we may see it with Pb but
not observe it for C, Ca, and Sn targets.
There also could be problems with the experimental 
data for Pb. As noted in \cite{Fissum96},
there is a systematic error in these data associated with the
correction for the electron contamination in the yield for the forward 
detectors with $Z \ge 20$ targets. For instance, because of the magnitude 
of this background, no experimental cross sections are reported for the 
Pb target at 51$^\circ$ \cite{Fissum96}.

\begin{figure}[h]

\begin{minipage}{8.0cm}
\vbox to 6.0cm {
\vspace*{-10mm}
\includegraphics[width=90mm,angle=-90]{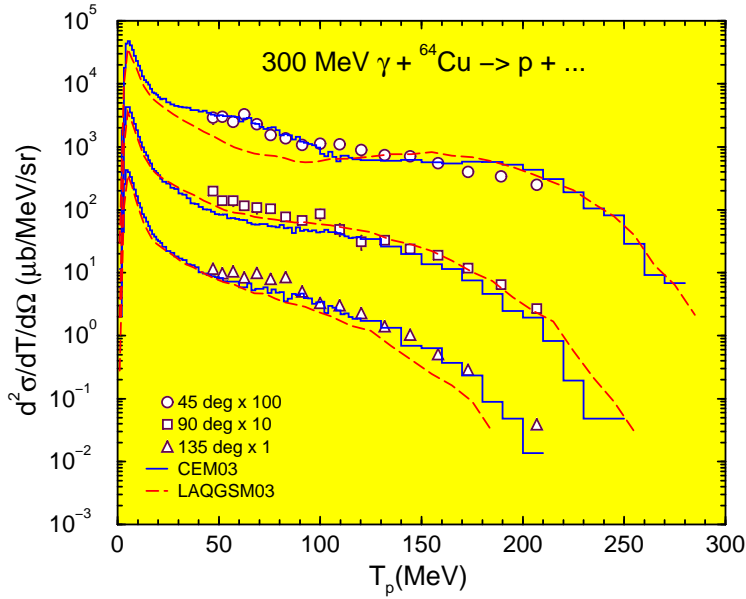}}
\end{minipage}
\hfill
\begin{minipage}{6.5cm}
\vspace*{-95mm}
\caption{
Proton spectra at 45$^{\circ}$, 90$^{\circ}$, and 135$^{\circ}$
from the reaction
300 MeV $\gamma$ + Cu. Symbols are experimental data from \cite{SCH82},
histograms and dashed lines are results of CEM03 and LAQGSM03,
respectively.
}
\end{minipage}

\vspace*{-65mm}
\end{figure}

\begin{figure}[h]

\begin{minipage}{8.0cm}
\vbox to 6.0cm {
\vspace*{-10mm}
\includegraphics[width=90mm,angle=-90]{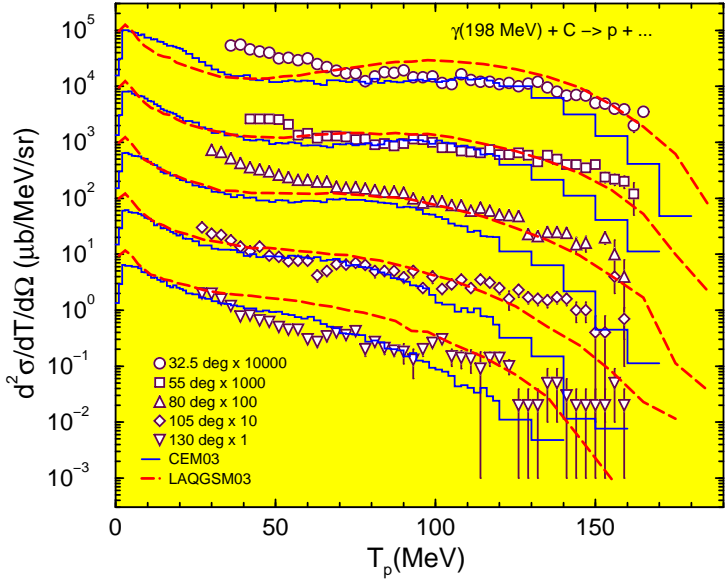}}
\end{minipage}
\hfill
\begin{minipage}{6.5cm}
\vspace*{-95mm}
\caption{
Proton spectra at 32.5$^{\circ}$, 55$^{\circ}$, 80$^{\circ}$, 
105$^{\circ}$, and 130$^{\circ}$
from the reaction
198 MeV $\gamma$ + C. Symbols are experimental data from \cite{ANG86},
histograms and dashed lines are results from CEM03 and LAQGSM03,
respectively.
}
\end{minipage}

\vspace*{-65mm}
\end{figure}

We now consider another type of photonuclear reaction,
induced by bremsstrahlung photons. In contrast to reactions
induced by monoenergetic photons of a given energy
$E$, the bremsstrahlung beam is
produced by monoenergetic electrons and has a spectrum
of photon energies $E$  of the form
$N(E,E_0) \sim 1/E$ \cite{Schiff51},
from 0 to $E_0$, where the end-point energy
$E_0$ is the maximum energy
of photons produced by the given electron beam. 
In addition, all experimental characteristics for
reactions induced by bremsstrahlung photons are
normalized per ``equivalent quanta", $Q$, defined as:
\begin{equation}
Q = {1\over E_0}\int\limits_0^{E_0} E \cdot N(E,E_0) dE \Biggm/
\int\limits_0^{E_0} N(E,E_0) dE 
\mbox{ .}
\end{equation}
As discussed above, since CEM03 and LAQGSM03 do not describe 
properly photonuclear reactions in the GDR region,  
we can calculate with our
codes bremsstrahlung reactions while limiting ourselves to
photon energies only above the GDR region.

\begin{figure}[h!]
\begin{minipage}{10.0cm}
\vbox to 6.0cm {
\vspace*{-10mm}
\hspace*{-2mm}
\includegraphics[width=77mm,angle=-90]{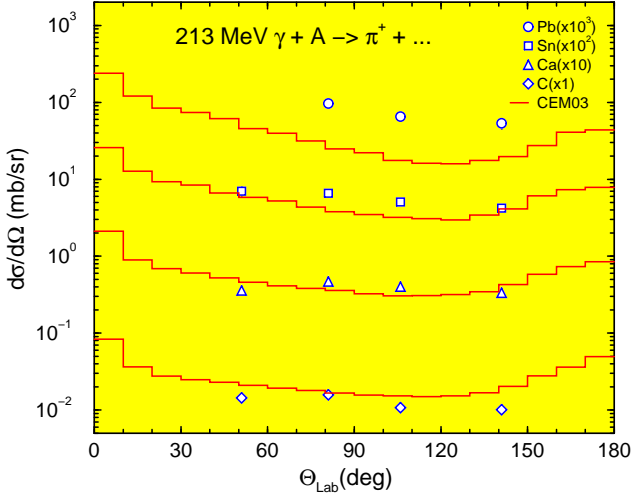}}
\end{minipage}
\hfill
\hspace*{-18mm}
\begin{minipage}{8.0cm}
\vspace*{-70mm}
\caption{
Energy-integrated angular distributions $d \sigma / d \Omega$ 
of $\pi^+$ emitted from 213 MeV $\gamma$ interactions with
Pb, Sn, Ca, and C. Symbols show experimental data by Fissum {\it et al.}
\cite{Fissum96} while histograms show CEM03 results. 
}
\end{minipage}
\vspace*{-70mm}
\end{figure}

This means we need to simulate in our calculations
the energies of the bremsstrahlung
photons according to their spectrum
$N(E,E_0) \sim 1/E$  up to $E_0$
not from 0, but from a value $E_{min}$,
above the GDR region, and in calculating the number of
equivalent quanta $Q$, we need to use $E_{min}$ for the lower
limits of the integrals in Eq.\ (46) instead of 0. This is easy to do
in our Monte Carlo calculations. After simulation 
of $N_{in}$ numbers of interactions of
bremsstrahlung gammas of energy $E_i$ with a nucleus,
the number of equivalent quanta $Q$ will be:
\begin{equation}
Q = {1\over {N_{in}E_0}} \sum_i E_i = {<E>\over E_0}
\mbox{ ,} 
\end{equation}
where the mean energy of the bresstrahlung photons $<E>$ is equal to
\begin{equation}
<E> \ = \ {{\int\limits_{E_{min}}^{E_0} E \cdot N(E,E_0) dE}
\over{\int\limits_{E_{min}}^{E_0} N(E,E_0) dE} }
=  {{\sum\limits_i E_i}\over {N_{in}}}\mbox{ ,} 
\end{equation}
and $E_{min} \leq E_i \leq E_0$. In the present paper, we use
$E_{min} = 30$ MeV for all the reactions we discuss.
The total inelastic (photoabsorption) cross section $\sigma_{in}$
in the case of bremsstrahlung photons is calculated as following:
\begin{equation}
\sigma_{in} = \ {{\int\limits_{E_{min}}^{E_0} \sigma^{\gamma}_{in}(E) 
\cdot N(E,E_0) dE}
\over{\int\limits_{E_{min}}^{E_0} N(E,E_0) dE} }
=  {{\sum\limits_i \sigma^{\gamma}_{in}(E_i)}\over {N_{in}}}\mbox{ ,} 
\end{equation}
where $\sigma^{\gamma}_{in}(E_i)$ is the photoabsorption cross section
by a nucleus of a photon with $E_{\gamma} = E_i$, simulated in a particular
Monte Carlo event $i$.

By using here for $E_{min}$ a value of 30 MeV instead of 0, we
will miss in our results the products 
from interaction of $\gamma$ with energies below 30 MeV,
like the cross sections for $(\gamma,n)$, $(\gamma,2n)$,
and, to a sertain degree,  $(\gamma,np)$, but this limitation
does not affect at all description of products in the deep spallation, 
fission (for preactinides),
and fragmentation regions, as well as the pion photoproduction
and spectra of nucleons and complex particles at energies
above the evaporation region.

Figs.\ 10 and 11 present examples of proton and $\pi^+$ spectra  
from bremsstrahlung interaction with carbon at $E_0 = 1050$
and 305 MeV, respectively. One can see that CEM03 describes
well both proton and pion measured spectra and agrees with 
the data better than the direct knockout model \cite{Boal81}
and quasideuteron calculations \cite{Matthews66} do.

\begin{figure}[h]

\begin{minipage}{8.0cm}
\vbox to 10.0cm {
\vspace*{-5mm}
\includegraphics[width=90mm,angle=-0]{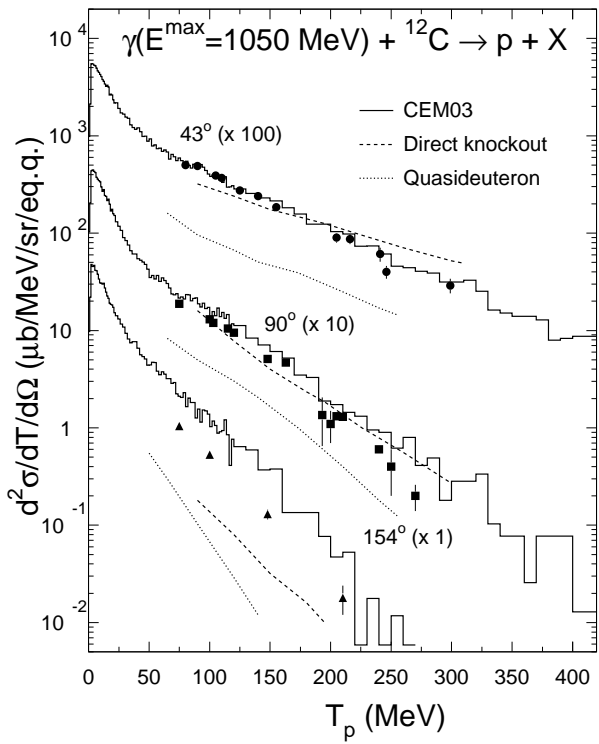}}
\end{minipage}
\hfill
\begin{minipage}{8.0cm}
\vspace*{-30mm}
\caption{
Comparison of measured \cite{Olson60}
differential cross section for proton photoproduction on
carbon at 43$^{\circ}$, 90$^{\circ}$, and 154$^{\circ}$ 
by bremsstrahlung photons with 
$E^{max} (\equiv E_0) = 1.05$ GeV
(symbols) with CEM03 calculations (histograms),
and predictions by the direct knockout model \cite{Boal81} (dashed lines)
and a quasideuteron calculation \cite{Matthews66} (dotted lines),
respectively.
The experimental data and results by the direct knockout and
quasideutron models were taken from Fig.\ 5 of Ref.\ \cite{Boal81}.
}
\end{minipage}

\end{figure}

\begin{figure}[h]

\begin{minipage}{7.0cm}
\vbox to 5.5cm {
\vspace*{-5mm}
\includegraphics[width=75mm,angle=-90]{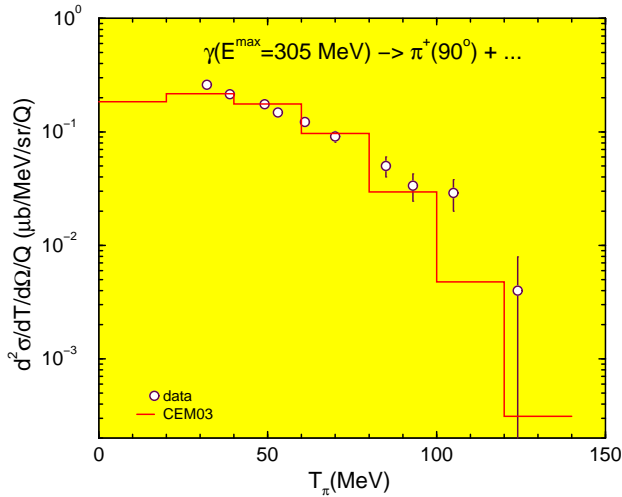}}
\end{minipage}
\hfill
\begin{minipage}{8.0cm}
\vspace*{-80mm}
\caption{
Comparison of measured \cite{Kabe64}
differential cross section for $\pi^+$ photoproduction on
carbon at 90$^{\circ}$ by bremsstrahlung photons with 
$E^{max} (\equiv E_0) = 305$ MeV
(circles) with CEM03 calculations (histogram).}
\end{minipage}
\vspace*{-60mm}
\end{figure}

Since the 1980's, a large number of radiochemical
measurements of bremsstrahlung-induced reactions have been performed
in Japan by the group of Prof. Koh Sakamoto
(see the recent reviews \cite{Sakamoto03,Haba02} and references therein).
Thousands of useful product cross sections were measured by
this group on target nuclei from $^7$Li to $^{209}$Bi
at bremsstrahlung end-point energies $E_0$ from 30 MeV to 1.2 GeV,
including photopion reactions, fragmentation and fission of preactinides,
deep spallation reactions, and recoil studies (mean kinetic energy and
the forward/backward (F/B) ratios of products). The authors of these
measurements have analyzed most of their data with the
PICA code by Tony Gabriel {\it et al.} \cite{Gabriel69,PICA},
with its improved version PICA95 \cite{PICA95,Sato99},
as well as with its latest version, PICA3, which
merged \cite{PICA3/GEM} with the mentioned above GEM
evaporation-fission code by Shiori Furihata \cite{GEM2},
{\it i.e.}, PICA3/GEM.         

Recently, this group provided us 
with numerical values of some of 
their measured reactions and we have calculated them with CEM03. 
Fig.\ 12 shows an example of the comparison of CEM03 results
with data for bremsstrahlung-induced 
fission cross sections of $^{197}$Au \cite{Haba00}
and $^{209}$Bi \cite{Haba02a}, compared as well with other available
experimental data for Au 
\cite{RAN67,JUN57}, \cite{Methasiri71}--\cite{Kiely76} and for Bi 
\cite{MOR69,RAN67,JUN57,Vartapetyan72,Emma76},
\cite{Bernardini53}--\cite{Carvalho75},
and with results by PICA3/GEM \cite{PICA3/GEM} from \cite{Haba02a}. 
There is a very good agreement of the CEM03 results with the
experimental data in the whole interval of $E_0$ measured, from
the threshold to the highest measured energy.

Fig.\ 13 presents experimental data
\cite{Haba00,Haba02a}, \cite{Sakamoto99}--\cite{Yamashita99}  
and calculations by PICA3/GEM  \cite{PICA3/GEM} and by CEM03
for the isotopic yields of products produced by bremsstrahlung
reactions on $^{197}$Au and $^{209}$Bi at $E_0 = 1$ GeV. 
For convenience, all the isotopes produced in these reactions were
divided into four groups, namely: 
  1) spallation products produced by sequental emission
of several nucleons, positive pions, and complex particles during the INC,
followed by preequilibrium and evaporation processes;
   2) intermediate-mass nuclides produced via fission of excited
compound nuclei;
   3) light fragments emitted either via evaporation or by ``fragmentation"
(Fermi break-up model, in the case of our present results), and
   4) ``photopion" products produced in $(\gamma,\pi^-xn)$ and
$(\gamma,2\pi^-xn)$ reactions, where the charge of the products
is higher than that of the initial target.
One can see that CEM03 describes the yields of products in
all these groups and agrees with the experimental data and results by
PICA3/GEM. CEM03 does not describe well the spallation
products very near the target, that are  
produced via $(\gamma,xn)$ reactions, because it does not consider 
photons with energies in the GDR region ($E_{min} = 30$ MeV), 
as discussed above. We note that the CEM03 and PICA3/GEM
results shown in the figure report $A$-distributions of the yield
of all products, {\it i.e.},
sums over $Z$ of yields of all isotopes with a given mass number $A$, 
while the experimental data 
obtained by the radiochemical method generally represent 
results for only several isotopes (sometimes, for only a single isotope) 
that contribute to the corresponding data point.
That is, this comparison is only qualitative but not quantitative and 
provides us only an approximate picture of the
agreement between the calculations and measured data.
Radiochemical measurements present the total yield for a given A
only for cases when cumulative cross sections that include contributions
from all precursors of all possible Z to the given measured yield;
therefore, in general theoretical calculations 
of A-distribution of yields should be higher than
many experimental radiochemical data points. 
A much better, quantitative analysis would be
to compare only the measured cross secitions,
isotope-by-isotope, as we did earlier for proton-induced reactions
(see, {\it e.g.}, \cite{GSI03,Titarenko02} and references therein).
We plan to perform such an analysis of isotopic yields from
photonuclear reactions in the future.

\newpage

\vspace{-5mm}
\begin{figure}[!ht]
\begin{center}

\includegraphics[width=14.0cm]{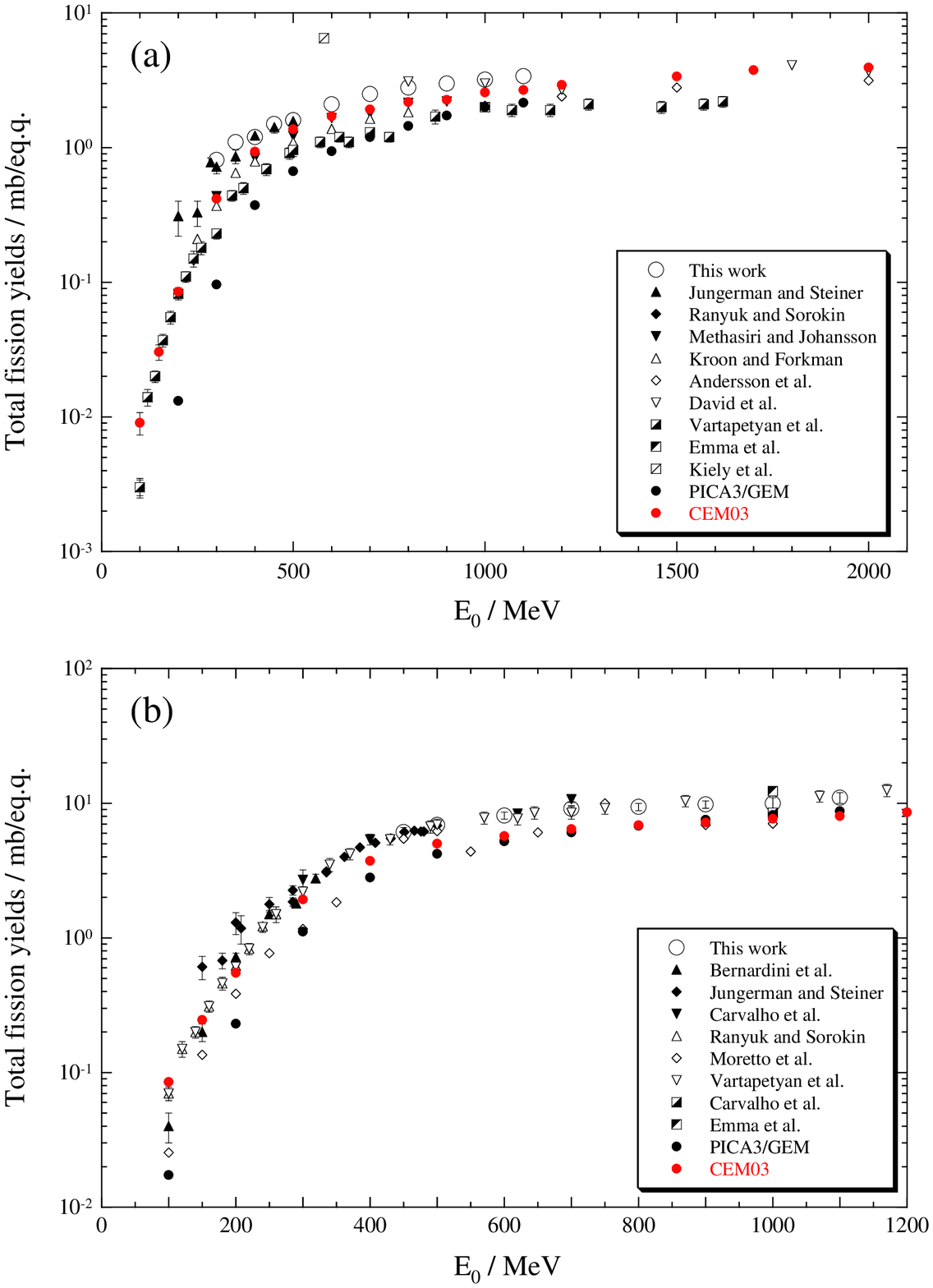}
\caption{
Bremsstrahlung-induced fission cross sections of $^{197}$Au (a) 
and $^{209}$Bi (b) as functions of the end-point energy $E_0$.
The experimental data for Au indicated in the insert of the
figure as ``This work" 
are from \cite{Haba00}; other experimental data on Au
are from \cite{RAN67,JUN57}, \cite{Methasiri71}--\cite{Kiely76},
as indicated. The data for Bi indicated as ``This work" 
are from \cite{Haba02a} and other data for Bi are from
\cite{MOR69,RAN67,JUN57,Vartapetyan72,Emma76},
\cite{Bernardini53}--\cite{Carvalho75}, as indicated.
The PICA3/GEM \cite{PICA3/GEM} results are from
\cite{Haba02a}; our present CEM03 results are shown as red circles.
We thank Dr. Hiromitsu Haba for making this figure for us
by adding our CEM03 results to Fig.\ 19 of the review \cite{ Sakamoto03}.
}
\end{center}
\end{figure}

\newpage
\vspace{-5mm}
\begin{figure}[!ht]
\begin{center}

\includegraphics[width=13.0cm]{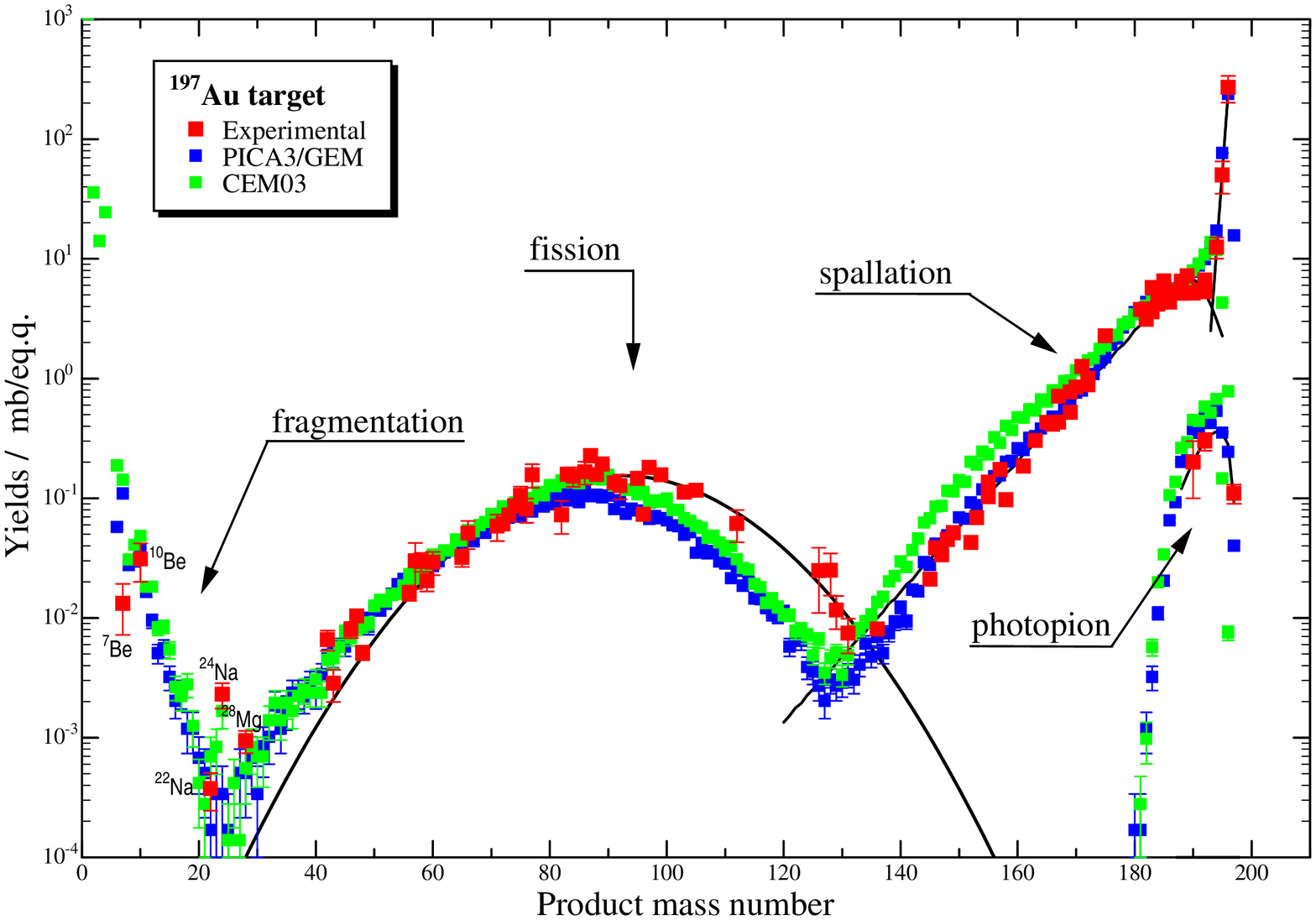}
\includegraphics[width=13.0cm]{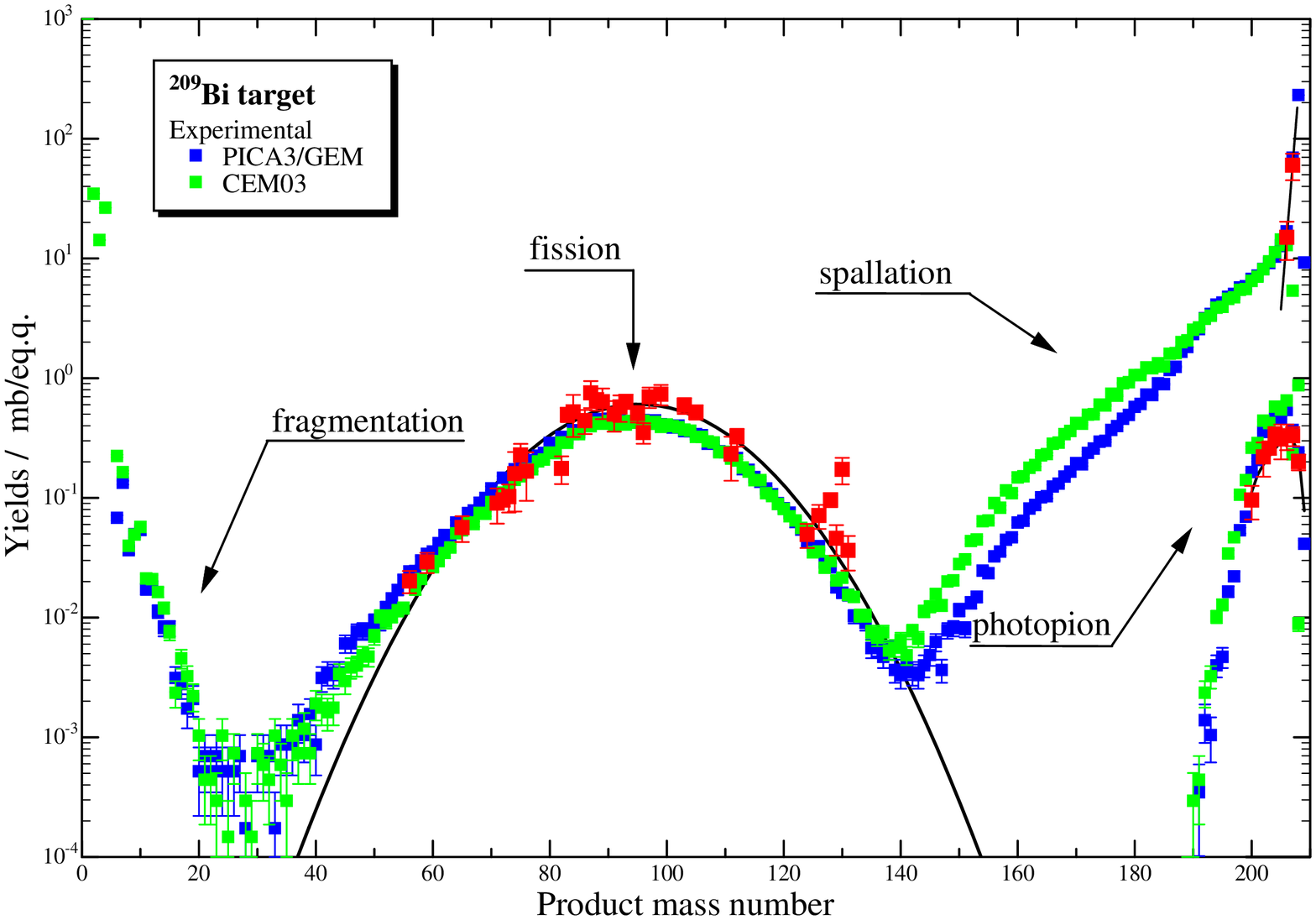}
\caption{
Comparison of CEM03 results 
for the isotopic yields of products produced by bremsstrahlung
reactions on $^{197}$Au and $^{209}$Bi at $E_0 = 1$ GeV
with experimental data \cite{Haba00,Haba02a}, 
\cite{Sakamoto99}--\cite{Yamashita99}  
and calculations by PICA3/GEM  \cite{PICA3/GEM}.
The experimental yields from fission of $^{197}$Au and $^{209}$Bi
are from Ref.\ \cite{Sakamoto99}; those by spallation on $^{197}$Au 
are from Refs.\ \cite{Sarkar91,Yamashita99}; those by fragmentation of
$^{197}$Au are from Ref.\ \cite{Matsumura00}; the PICA3/GEM results
are from several publications of Prof.\ Sakamoto's group and
are presented in Fig.\ 18 of the review \cite{Sakamoto03} with the
corresponding citations. The mass yields for the fission
products shown by black curves represent approximations based on 
experimental data obtained in Refs.\ \cite{Haba00,Haba02a}.
We thank Dr.\ Hiroshi Matsumura for making this figure for us
by adding our CEM03 results to Fig.\ 18 of the review \cite{Sakamoto03}.
}
\end{center}
\end{figure}
      
Our preliminary analysis shows that CEM03 also allows us to describe 
the recoil properties
(forward and backward product yields, their F/B ratio, and mean kinetic
energies) of nuclides produced in bremsstrahlung-induced
reactions on medium and heavy targets at intermediate energies  
(see \cite{Haba02} and reference therein). We plan to publish
our analysis in a future paper. Here we present only 
several predictions by CEM03 for the reaction of $E_0 = 1$ GeV 
bresstrahlung photons on Au, as we find such results 
informative and useful to
better understand the mechanisms of nuclear reactions.

Due to the momentum transferred by the bombarding gammas to the
nuclear target, one may expect that most of the spallation products
would fly in the forward direction in the laboratory system.
The lower-right plot in Fig.\ 14 shows the mean laboratory angle
$\Theta$ of all products as a function of $A$. We see that
the mean angle of most spallation products is predicted by CEM03
to be between 72 and 80 degrees. (It is not equal to 0 degrees, as the
probability of projectiles to have an impact parameter exactly
equal to zero is equal to zero, and photons hit more often the
periphery of the nucleus rather than its center.) 
The black solid curve in the upper-left plot of
Fig.\ 14 shows 
the yield of all products as a function of $A$, the same results compared
in Fig.\ 13 with experimerntal data and calculations by PICA3/GEM.
Besides the total yield, this plot shows also its components from 
nuclides produced in the forward (long-dashed red line)
and backward (blue dashed line) directions. 
One can see that for all the spallation isotopes,
the cross sections for the forward products are about a factor of two
higher than for backward products, in complete accordance 
with the available experimental data (see the 
review \cite{Haba02} and references therein).
But the situation changs completely for fission products:
The momenta of the fissioning nuclei is small,
their mean kinetic energy in the laboratory system is
a few MeV (see the upper-right plot in Fig.\ 14), that is much
less than the kinetic energy from several tens
to about a hundered  MeV that fission fragments receive due to the 
Coulomb repulsion of the fragments.
If we neglect the effects of angular momentum, the fission fragments 
would be distributed isotropically
in the system of the fissioning nucleus, and the small momentum of
the fissioning nuclei makes this distribution
almost isotropic also in the laboratory
system. The upper left plot of Fig.\ 14 shows that predicted
yields for the fission fragments in the forward and backward directions
are almost the same, {\it i.e.}, the F/B ratio for the fission
fragments is almost equal to one, again in complete agreement
with the available experimental data
(see \cite{Haba02} and references therein).

The mean kinetic energy of the forward products shown in the upper-right
plot of Fig.\ 14 is only
very slightly higher than that of the backward products
(the momenta of fissioning nuclei are low, as discussed above), with
a little higher effect for the spallation products than for fission
fragments, as is to be expected. Due to this fact and to the near
isotropy of the fission fragments, 
some fission fragments may have
their mean velocity in the direction opposite the beam,
as can be seen from in the lower-left plot of this figure.

It is much more informative to study the F/B problem considering 
the forward and backward cross sections
for every product separately, as shown in Figs.\ 15 and 16, rather than
addressing only the A-distribution of their yields. Whereas the Z-averaged
A-dependence of the F/B ratio is about a factor of two for all the
spallation region (see also Fig.\ 17), 
the situation changes for individual isotopes.
The cross sections of forward-emitted isotopes are still about a factor
of two higher than the backward cross sections
for most of the spallation products, but their ratio 
is much higher for Ho and Rh, and depends strongly on the mass numbers
of the products. Ho and Rh
are ``photopion" products produced via $(\gamma,\pi^-xn)$ and 
$(\gamma,2\pi^-xn)$, respectively, with emisssion of only a few
neutrons in addition to the pions. When the number of emitted
neutrons is small, the product ``remembers" the momentum
transfered to the target by the projectile, and such neutron-rich
products go mainly forward, with a ratio F/B up to ten 
or higher. On the contrary, in reactions where many neutrons are 
emitted approximately isotropically, the residual nucleus has lost
most of its ``memory" of the initial momentum.
Therefore the neutron-deficient products from such reactions 
have a smaller F/B ratio, 

\vspace{-5mm}
\begin{figure}[!ht]
\begin{center}

\includegraphics[width=15.5cm]{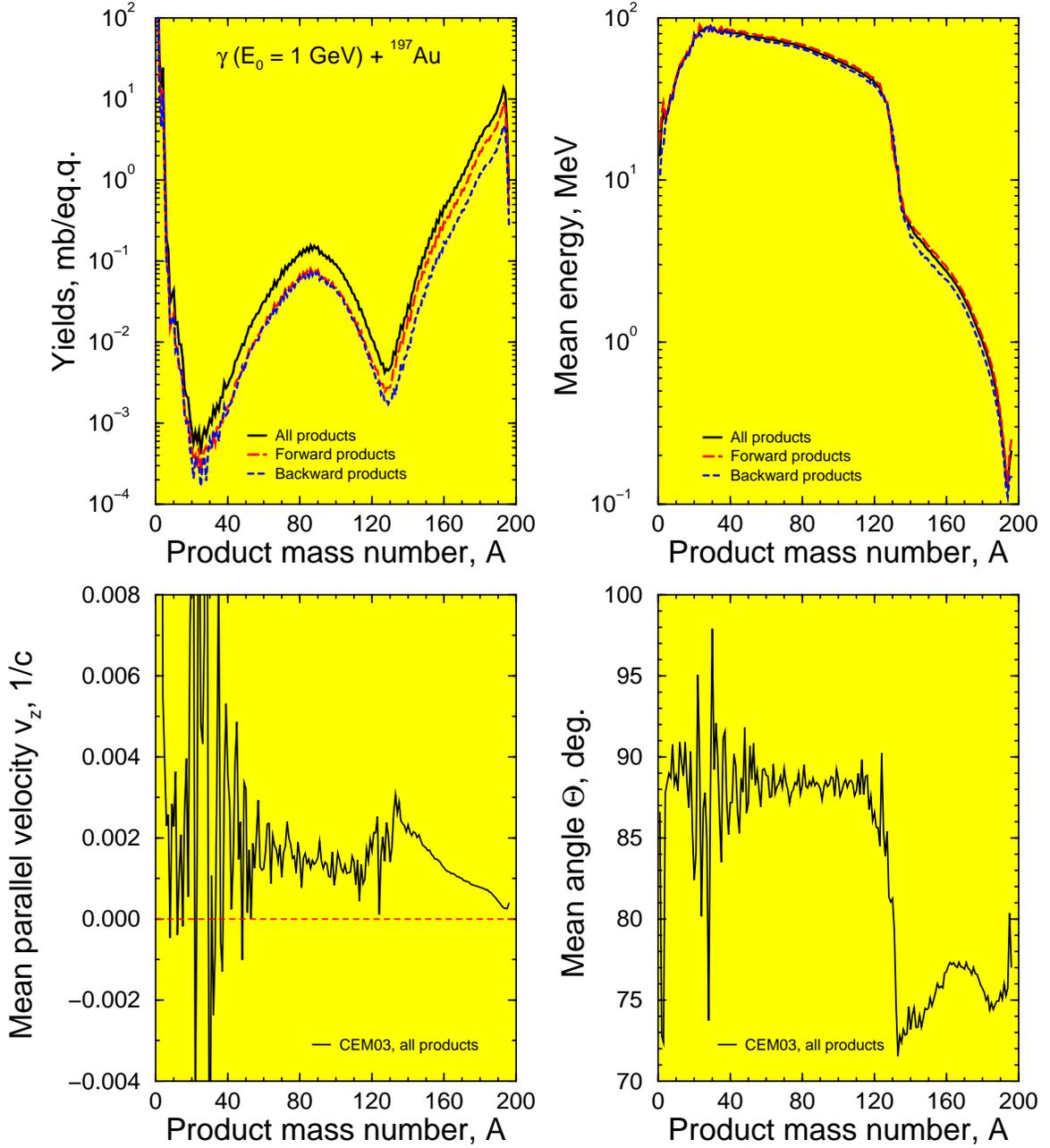}
\caption{
Results by CEM03 for 1 GeV bremsstrahlung-induced reactions
on Au.
{\bf Upper left plot:} mass yield of all products (black line),
isotopes poduced in the forward laboratory direction 
(long-dashed red line), and backward products (dashed blue line);
{\bf Upper right plot:} mean laboratory kinetic energy of all products
(black line) and of only forward (long-dashed red line) and 
 backward products (dashed blue line);
{\bf Lower left plot:} mean laboratory velocity $v_z$ of all 
products in the beam direction;
{\bf Lower right plot:} mean laboratory angle $\Theta$ of
all products as a function of $A$. The big fluctuations in the
values of $v_z$ and $\Theta$ for masses around $A = 20$ and
130 do not provide real physical information,
as they are related to the limited statistics of our 
Monte Carlo simulation caused by the very low yield of isotopes 
at the border between spallation 
and fission, and at that between fission and fragmentation.
Our calculation provides only a few (or even one) 
isotopes of a given $A$ in these mass regions, and mean values
for such events do not have any significance.
}
\end{center}
\end{figure}
\newpage

\newpage
\begin{figure}[!ht]
\begin{center}
\vspace{-10mm}
\includegraphics[width=16.0cm]{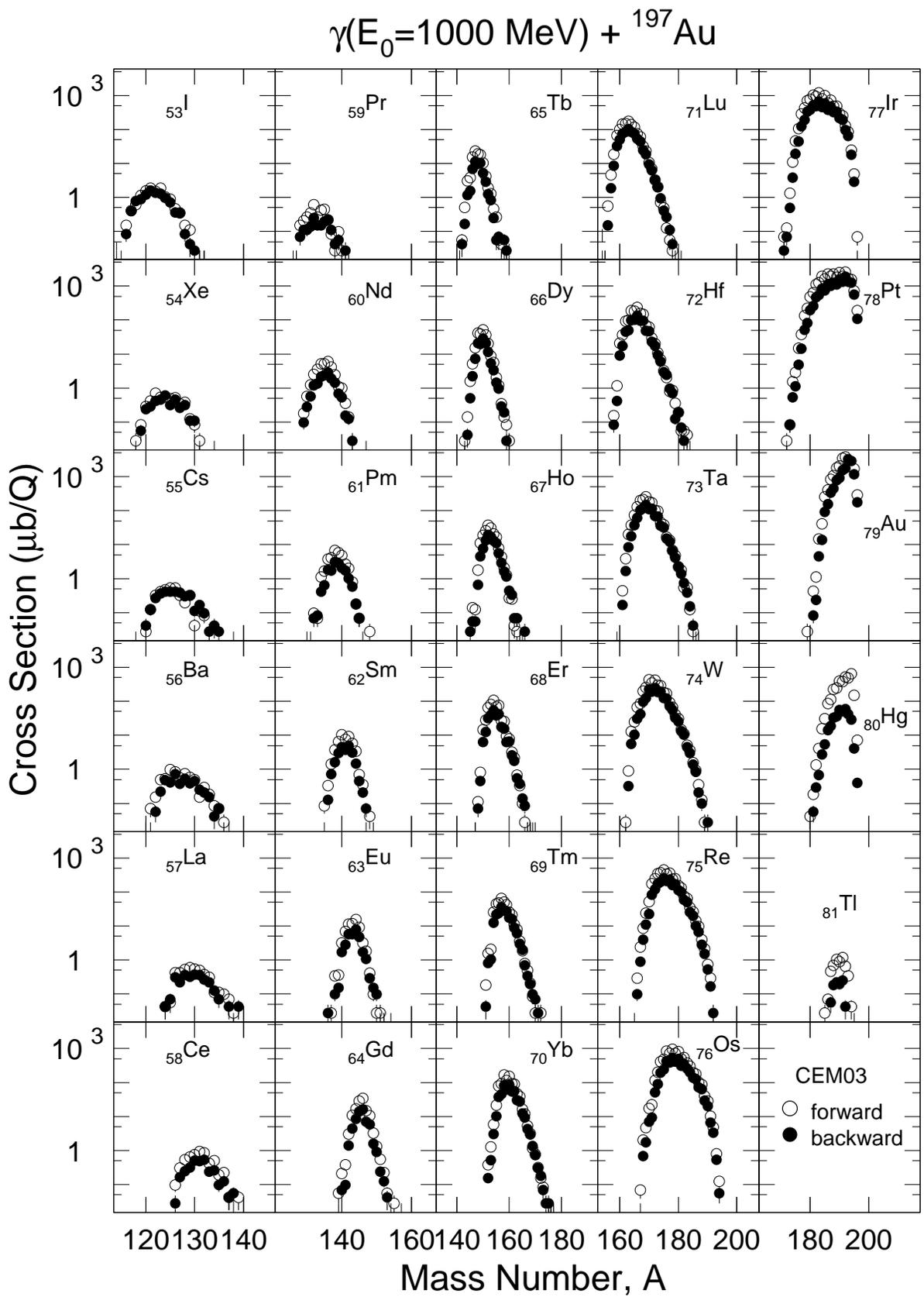}
\caption{
Predicted cross sections of the spallation products
from the $E_0 = 1$ GeV bremsstrahlung photon-induced reaction on Au: 
Open circles
show the yield of the products produced in the forward direction 
in the laboratory system, while the black circles show 
results for backward products.
}
\end{center}
\end{figure}

\newpage
\begin{figure}[!ht]
\begin{center}
\vspace{-10mm}
\includegraphics[width=16.0cm]{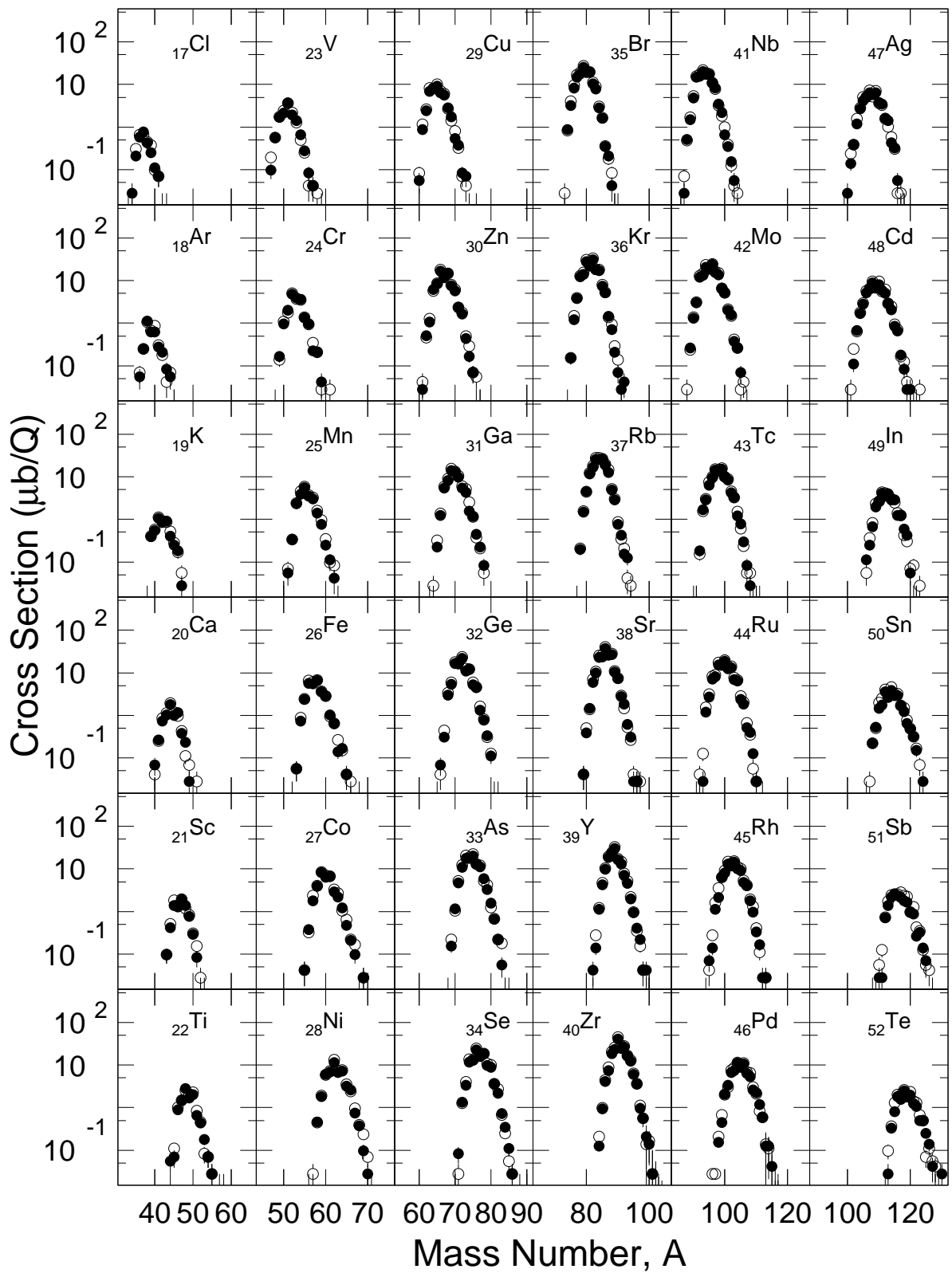}
\caption{
The same as Fig.\ 15, but for fission and fragmentation products. 
}
\end{center}
\end{figure}
\newpage

\begin{figure}[!h]

\begin{minipage}{7.0cm}
\vbox to 10.5cm {
\includegraphics[width=85mm,angle=-0]{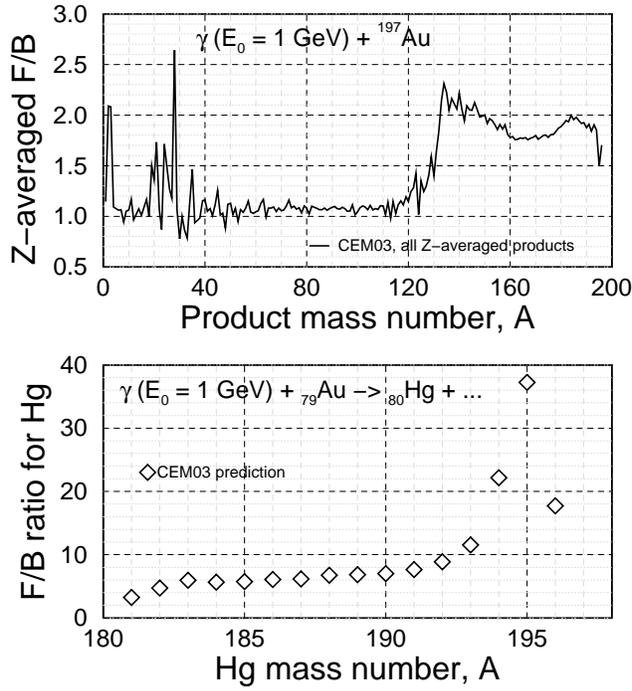}}
\end{minipage}
\hfill
\begin{minipage}{8.0cm}
\vspace*{-60mm}
\caption{
{\bf Upper plot:}
Z-averaged A-dependence of the F/B ratio of the 
forward product cross sections to the backward ones,
as predicted by CEM03 for the reaction of
$E_0 = 1$ GeV bresstrahlung photons on Au.
{\bf Lower plot:} The F/B ratio of the Hg isotopes
produced in the same reaction as a function of the
mass number of the Hg nuclei.
}
\end{minipage}

\end{figure}

{\noindent
usually around a factor of two.}
The farther away from the target are the products, the smaller
is this effect; for products with $Z \lesssim 70$, it practically 
disappears. Approaching the border of the transition between spallation
and fission products, the F/B ratio decreases and for Ce, La, Ba, Cs,
Xe, and I nuclei shown in the left column of Fig.\ 15, 
the F/B ratio becomes almost equal to one and remains so
for all the fission products shown in Fig.\ 16.

The $Z$-averaged F/B ratio for all nuclides of a given $A$ as a function of A
is presented in the upper plot of Fig.\ 17. The lower plot of this figure
show the F/B ration for isotopes of Hg: One can see that it decreases
from about thirty-seven for neutron-rich $^{195}$Hg to about three for 
neutron-deficient $^{181}$Hg.

We think that analysis of such recoil characteristics is quite informative
not only for photonuclear reactions, but also for proton-induced
and other types of reactions. Analysis of experimental data for such
characteristics would allow us to understand better the mechanisms of nuclear
reactions and may help us to distinguish 
the fission processes from the fragmentation (or evaporation) ones
in production of heavy fragments from reactions on medium-mass targets, 
like Fe (see disussion of this problem in \cite{OurNewINC,Madeira04}).
New measurements on the recoil properties from reactions
with any type of projectiles, including bremsstrahlung photons,
would be very useful.

{\noindent {\bf Conclusions}}

The 2003 versions of the codes CEM2k+GEM2 and LAQGSM, CEM03 and LAQGSM03,
are extended to describe photonuclear reactions. Both our models
consider photoproduction of at most two pions, which limits their 
reliable application to photon energies up to only about 1.5 GeV.
The present version of our models do not consider photoabsorption
in the GDR region, which defines the lower limit of the photon energy
to about 30 MeV. 
Nevertheless, developing and incorporating into CEM03 a routine based on
the phenomenological systematics for the total photoabsorption cross section
by Kossov allow us to enlarge the region of applicability of CEM03
and to get quite reasonable results for applications both in the GDR 
region and above 1.5 GeV.

As shown by several examples, CEM03 and LAQGSM03 allow us
to describe reasonably well, and better than with their precursors,
many photonuclear reactions 
needed for applications, as well as to analyze 
mechanisms of photonuclear reactions for fundamental
studies.
But our models still have several problems. 
Fig.\ 18 shows examples of such problems on 
proton and deuteron spectra from reactions
induced by 60 MeV monoenergetic photons on Ca:
One can see that both CEM03 and LAQGSM03 desrcibe reasonably well
the shape of the proton spectrum, but their absolute values differ by more
than a factor of two. This is because the CEM03 results are normalized to the
total photoabsorption cross section predicted by the Kossov systematics,
which gives 3.49 mb for this reaction, while the LAQGSM03 results are
normalized to the Monte-Carlo-calculated
total photoabsorption cross section of 8.5 mb. If we refer to Fig.\ 4,
we see that the Kossov systematics for the reaction $\gamma + $ Ca
predict values that are a factor of two below available experimental data
at energies around 60 MeV. This explains the difference we get
between the CEM03 and LAQGSM03 results shown in Fig.\ 18, and suggest
that the Kossov systematics should be further improved.  Even allowing
for this normalization problem, both codes appear to underestimate
the cross sections for the higher-energy protons.

A further problem shown in Fig.\ 18 is for the spectra of deuterons.
The predicted spectra of deuterons differ both in their shapes and 
absolute values for the two codes.  CEM03 and LAQGSM03 have different 
intranuclear-cascade
models, leading after the INC stage of any reaction to
different average values for $A$, $Z$, and $E$ of the excited nuclei, 
as a starting point for the preequilibrium and evaporation stages 
of reactions,
where most of the deuterons are produced. This explains the 
difference in the deuteron spectra predicted by the two codes
and suggests that further work to improve the description of 
complex-particle emission is necessary.

\begin{figure}[!h]

\begin{minipage}{7.0cm}
\vbox to 6.5cm {
\vspace*{-10mm}
\includegraphics[width=75mm,angle=-90]{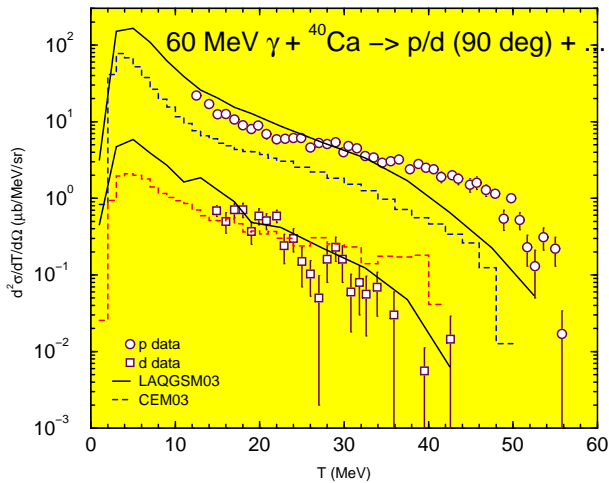}}
\end{minipage}
\hfill
\begin{minipage}{8.0cm}
\vspace*{-80mm}
\caption{
Examples of two problems with the current versions
of our codes: proton and deuteron spectra at 90$^{\circ}$
from the reaction
60 MeV $\gamma$ + Ca. Symbols are experimental data from \cite{RYC90},
solid lines and dashed histograms are results from LAQGSM03 and CEM03,
respectively. The CEM03 spectra are normalized to the total
absorption cross setion as predicted by Kossov's systematics,
equal to 3.49 mb for this reaction, while the LAQGSM03 spectra
are normalized to the Monte-Carlo total absorption cross section 
calculated by that code to be equal to 8.5 mb. 
}
\end{minipage}

\vspace*{-70mm}
\end{figure}

The overestimation of the high-energy
tail of the deuteron spectrum by CEM03 is partially related with
an imperfect description of the preequilibrium emission
of $d$ from this reaction, due to an excessively simplified
estimation of the probability of several excited
nucleons (exitons) coalescing into a complex particle
that can be emitted during the decay of the excited
nuclei produced after the cascade.

We plan to address these problem in the future. In addition,
we plan to extend our models to describe photoabsorption in the
GDR region, as discussed previously, and to extend
our models to describe photonuclear reactions at energies
of 10 GeV or more.

Our present study suggests that analyzing characteristics of recoil
nuclei produced by photonuclear and other types of reactions is a
powerfool tool to understand mechanisms of nuclear reactions.
We encourage future measurements of such characteristics
both for photonuclear and proton- or/and nucleus-induced reactions.

{\noindent {\bf Acknowledgements}}

\indent
We thank Dr.\ Kumataro Ukai for providing us with numerical values of single
pion photoproduction cross sections from their compilation \cite{UKA97}.
We are grateful to Dr.\ Igor Pshenichnov for sending us the $\gamma p$
and $\gamma n$ event generators used in their Moscow photonuclear
reaction INC \cite{Iljinov97}. 
We thank Prof.\ Koh Sakamoto and Drs.\ Hiroshi Matsumura, Hiromitsu Haba,
and Yasuji Oura for providing us with their publications and numerical
tables of their measured data, as well as for useful discussions, help
in drawing several figures for us, and their interest in our modeling.

This work was partially supported by the US Department of Energy,
the Moldovan-US Bilateral Grants Program, CRDF Project MP2-3045-CH-02,
and the NASA ATP01 Grant NRA-01-01-ATP-066.

\end{document}